\documentclass[aps,pre,twocolumn,showpacs,groupedaddress,footinbib]{revtex4-1}
\usepackage{natbib,hyperref}
\hypersetup{colorlinks=true, citecolor=blue, urlcolor=blue, linkcolor=blue}
\usepackage[english]{babel}
\usepackage{graphicx}
\usepackage[caption=false]{subfig}
\usepackage{amsmath}
\usepackage{amsfonts}
\usepackage{amssymb}
\usepackage{color,soul}
\setulcolor{red}
\usepackage{xcolor}
\usepackage[utf8]{inputenc}
\begin{document}

\title{Emergence of rotating clusters  in active Brownian particles with visual perception}

\author{Radha Madhab Chandra}
\thanks{These two authors contributed equally} 
\author{Alan Biju John}
\thanks{These two authors contributed equally} 
\author{A. V. Anil Kumar}
\email{anil@niser.ac.in}
\affiliation{School of Physical Sciences, 
National Institute of Science Education and Research, Jatni, Bhubaneswar 752050, India.}
\affiliation{Homi Bhabha National Institute, AnushakthiNagar, Mumbai}

\begin{abstract}

  We examine the group formation and subsequent dynamics of active particles which are equipped with a visual perception using Langevin dynamics simulations. These particles possess an orientational response to the position of the nearest neighbours which are within a vision cone of these particles. We observe the emergence of rotating clusters when the visual perception of the particles are in the intermediate range.  We have  found that the persistent motion of these active particles are intimately correlated with the emerging structures by analysing the persistence probability as well as the orientational correlation function.  For rotating clusters,  the persistent probability is found to be very quickly decaying and orientational correlation function shows oscillatory behaviour.  
 
\end{abstract}

\maketitle

\section{Introduction}

Active systems are characterised by their ability to self propel at the expense of energy consumption from the environment or created by an internal mechanism\cite{ramaswamy1,vicsek1,elgeti}. These systems are ubiquitous in nature and can be found in different scales ranging from active colloids to biological cells to higher level organisms like fish and birds. The self propulsion and resultant dynamic phenomena have exalted much interest with the formation of a distinct class in non-equilibrium statistical physics that is intrinsically out of thermal equilibrium.  Moreover these systems break time reversal symmetry due to continuous energy dissipation of individual constituents. 

One of the astounding features exhibited by active matter systems is the the collective, coherent motion of large numbers of organisms like flocking of birds, school of fish, ant milling, herd of mammals etc. Such collective behaviours have been studied earlier with the Boids model\cite{reynolds} introduced by Reynolds in 1987, the pioneering Vicsek model proposed in 1995\cite{vicsek} as well as the behavioural zonal model introduced by Couzin in 2002\cite{couzin, couzin1}. This spontaneous, synchronized motion, omnipresent in nature spans an enormous range of length scales from micrometers (e.g,the micro-organism Dictyostelium discoideum) to metres and is characterized by alignment, attraction and avoidance(“the three A’s”) which is indispensable for their ecological survival (finding food, avoiding predators) too. Such flocking organisms possess nonzero center of mass velocity for the flock as a whole and thus spontaneously breaking the rotational symmetry. The non-equilibrium nature of active matter allows flocks to achieve long-range order even in two dimensions\cite{toner}, escaping the Mermin-Wagner-Hohenberg theorem constraints\cite{mermin,hohenberg}. Moreover the seasonal migrations of birds and mammals clearly signifies that they also do not move in a rotational-invariant environment. 

Many other interesting behaviours have been predicted for active systems in addition to flocking such as giant density fluctuations\cite{ramaswamy, narayan} and spontaneous phase separation\cite{tailleur, redner}. When a large number of active particles interacting isotropically with each other ands move sufficiently rapidly, they come together and form large clusters even in the absence of any attractive interactions, leading to a phenomenon called motility induced phase separation(MIPS)\cite{filly, cates}. Most of these interesting properties of active systems can be obtained from simple models which do not take into account of the environment explicitly. For example, Vicsek model\cite{vicsek1} employs an alignment interaction between active particle and its neighbours in addition to the intrinsic noise to predict a continuous transition from a disordered to flocking state. However, an explicit velocity alignment interaction is not a necessary condition for flocking. Flocking can be achieved by purely local physical interaction\cite{grossman,strombom,szabo,peruani}. A more generic model of active particles is the active Brownian particles(ABPs) in which active particles are modelled as agents moving with a constant velocity whose direction will undergo changes randomly with Gaussian delta-correlated noise term incorporating the rotational diffusion of these particles as well as the interaction between them. In this model, active particles with purely repulsive interactions exhibit many of the features outlined above, such as MIPS\cite{redner, filly,cates}, wall accumulation\cite{elgeti1} etc. 
The exploration of these models and different  phenomena observed in them provides essential insights on the development of artificial systems that emulate such behaviours such as self-propelled colloidal particles including magnetic-bead-based colloids that replicate artificial flagella\cite{dreyfus}, catalytic Janus particles\cite{howse, erbe, palacci, baraban}, platinum-loaded stomatocytes\cite{wilson}.

In many active systems, the activity actually depends on the local  environment such as density, visual perception etc., leading to nonreciprocal interactions. For e.g., many bacteria use autoinducer molecules to sense the local density and modify their gene expression to regulate the virulence. This phenomenon is known as quorum sensing\cite{mukherjee,miller}.  Here the particles go through temporal active-passive switching depending on the local environment which leads to many interesting phenomena such as glass to demixed phase to liquid\cite{jalim}, cluster formation\cite{bauerle},  oscillatory colloidal waves\cite{cao} to name a few. 

Another interesting nonreciprocal mechanism is visual perception of active particles, where they vary their active velocity based on the external feedback from their immediate neighbours\cite{lavergne}. It has been shown that new structures emerges  when the active particles are equipped with a visual perception\cite{barberis,negi}.  In this model, the active brownian particles detects their immediate neighbours within a vision cone and modify the direction of active velocity accordingly.  These intelligent active brownian particles(iABPs) exhibit different phases such as dilute fluids, worms, worms-aggregates coexistence and hexagonal aggregates as we vary the vision angle of these particles.   We have carried out Langevin dynamics simulations in this model system to study the collective motion of aggregates and their dependence on the extent of visual perception. We have observed that rotating clusters emerge when the vision angle of the active particles is in the intermediate range. Such spontaneously emerging rotating clusters are observed earlier in simulations\cite{yang, huang, cantisan} as well as in experiments\cite{kokot}.  This rotating clusters are mainly found in the chiral active particles where the particles are asymmetric due to mass distribution, surface coating or body shape. Also, in many of these situations rotating clusters are observed when the motion is deterministic\cite{ cantisan}. However, we found that the rotating clusters emerge in our study even when the stochastic terms are present in the equations of motion. We have attempted to correlate this rotational motion of the clusters with the persistent motion of the active particles by calculating the persistence probability and orientational correlation function.  We found that the persistent motion is less when the aggregates are larger in size  and these larger clusters rotates as a whole. 

The remainder of this article is organised as follows. In section II, we detail our model and the interaction potentials. The simulation details are also outlined in this section. Our results are presented in section III along with discussions. We finally summarise the main results in section IV.

\section{Modeling and simulation framework}

In our model we have considered $N$ sensitive as well as responsive “intelligent" active Brownian particles(iABPs) in two dimensions at positions $\mathbf{r}_{i}(t)$ ($i = 1,. . ., N$) at time $t$, which are propelled with constant velocity along its orientation vector which evolves stochastically. The translational motion is governed by underdamped Langevin equations which are given by
\begin{equation}\label{1}
    m_i \ddot{\mathbf{r}}_i = - \gamma \dot{\mathbf{r}}_i + \sum_{ij} \mathbf{F}_{ij} +  \gamma v_{0} \boldsymbol{e}_{i}(t) + \sqrt{2m\gamma k_B T } \, \boldsymbol{\eta}_i(t)
\end{equation}
where the acceleration term $m\boldsymbol{\ddot{r}}(t)$ accounts for translational inertia. The total force on the right-hand-side of equation is given by the sum of the frictional force $-\gamma\boldsymbol{\dot{r}}(t)$,
proportional to the translational friction coefficient and a conservative force $\boldsymbol{F_{ij}} = -\nabla_{i}\boldsymbol{V_{ij}}(r_{ij}(t))$ which describes excluded volume interactions between the iABPs. The interaction between the particles are modelled via a purely repulsive, short-ranged, truncated and shifted Lennard-jones Potential or Weeks–Chandler–Andersen (WCA) potential given by
\begin{equation}\label{2}
    V(r_{ij}) = 4 \epsilon \left[ \left( \frac{\sigma_{ij}}{r_{ij}} \right)^{12} - \left( \frac{\sigma_{ij}}{r_{ij}} \right)^{6} + \frac{1}{4} \right] \Theta({2^\frac{1}{6}}\sigma_{ij} - r_{ij}    )
\end{equation}
where $r_{ij} = \left| \vec{r}_j - \vec{r}_i \right|$ is the distance between particles $i$ and $j$ and $\epsilon$ is the interaction strength. Here, $\sigma_{ij} = (\sigma_{i} + \sigma_{j})/2$ is the mean diameter of the two particles with $\sigma_{i}$ being the diameter of particle $i$. We consider one-component system here thus $\sigma_{ij} = \sigma$. The Heaviside step function, $\Theta({2^\frac{1}{6}}\sigma_{ij} - r_{ij})$, implements a cutoff of the potential at $2^\frac{1}{6}\sigma$. A stochastic thermal force ${\eta_{i}}$, characterized by Gaussian and Markovian white noise of zero mean and unit variance with both spatially and temporally delta-correlated, i.e $\langle\eta_{i}\rangle = 0 , \langle \eta_{i\alpha}(t) \eta_{j\beta}(t') \rangle = \delta_{ij} \delta_{\alpha \beta} \delta(t - t')$ describes the fluctuations or random collisions with the solvent molecules. The intensity of the noise is related with the translational diffusion coefficient as $D_{T} = \frac{k_{B}T}{\gamma}$. The self propulsion of the particles is described by
$\gamma v_{0} \boldsymbol{{e}_i}$, where $v_0$ is magnitude of the active velocity, which is kept constant throughout the simulations.
This active force $\gamma v_{0} \boldsymbol{{e}_i}$ in Eq. (1) couples the translational motion to the rotational motion of the particles via the orientation vector i.e $ \boldsymbol{e_{i}} = (cos \phi_{i}, sin \phi_{i})^{T} $. In normal active brownian particles, the direction of the self-propulsion velocity vector $e_{i}(t)$ undergoes free Brownian rotation according to 
\begin{equation}\label{3}
    \boldsymbol{\dot{e}}_i(t) = \sqrt{2D_r} \left( \boldsymbol{\xi}(t) \times  \boldsymbol{e}_i(t) \right)
\end{equation}

\noindent where $D_{r}$ accounts for the rotational diffusion coefficient. $\boldsymbol{\xi}(t)$ represents random torque characterized by Gaussian and Markovian noise with zero mean and correlation $\langle \boldsymbol{\xi}_i(t) . {\boldsymbol{\xi}}_j(t') \rangle = \delta_{ij} \delta(t - t')$.
 Since an iABP is responsive to its neighbours in its vision cone, particle $i$ at position $r_{i}$  adjusts its propulsion direction $\boldsymbol{{e}}_i$ through self steering towards the $j$ th particle in the direction $\hat{\mathbf{r}}_{ij} = \frac{\mathbf{r}_j - \mathbf{r}_i}{|\mathbf{r}_j - \mathbf{r}_i|}$, with an adaptive torque as already have been seen in cognitive flocking model\cite{negi}. Thus the complete equation for the stochastic evolution of orientation vector is given as :
 \begin{equation}
    \boldsymbol{\dot{e}}_i(t) = \sqrt{2D_r} \left( \boldsymbol{\xi}(t) \times  \boldsymbol{e}_i(t) \right) + \frac{\Omega_v}{N_{c,i}} \sum_{j \in VC} e^{-r_{ij}/R_0} \mathbf{e}_i \times (\hat{\mathbf{r}}_{ij} \times \mathbf{e}_i)
    \label{label4}
\end{equation}
For our two-dimensional system in terms of orientation angle $\phi_{i}$ the equation \eqref{label4} reduces as\cite{negi} 
\begin{equation} \label{label04}
    \dot{\phi}_i = \frac{\Omega_v}{N_{c,i}} \sum_{j \in VC} e^{-r_{ij}/R_0} \sin(\phi_{ij} - \phi_i) + \sqrt{2D_r} \xi_i(t)
\end{equation} 
$\Omega_v$ is known as visual maneuverability, i.e how quickly a particle can react to external stimuli by adjusting its orientation and thus becoming more  maneuverable, $\phi_{ij}$ is the polar angle of unit vector $\hat{\mathbf{r}}_{ij}$, and $R_{0}$ is the characteristic length. The decaying exponential prefactor of distance signifies that an iABP will be more responsive towards the nearby particles rather than the distant ones. The effective number of particles within vision cone of reference particle $i$ is given as
\begin{equation}
   N_{c,i} =  \sum_{j \in VC} e^{-r_{ij}/R_0} 
\end{equation}
Such exponential decaying form considers the partial blocking off the vision range by nearby particles.
As mentioned in \cite{negi} the condition for particles $j$ lies
within the "retina" of particle $i$ is
\begin{equation}
    \mathbf{\hat{r}}_{ij} \cdot \mathbf{e}_i \geq \cos(\theta)
\end{equation}where $\theta$ is half of
the opening angle of the vision cone centred by the particle orientation $e_{i}$. In addition we will consider visual range upto ${|\mathbf{r}_j - \mathbf{r}_i|} \leq 4R_{0}$ and treat others as invisible. The details of this model can be found in Ref.\cite{negi}.  The equations of motion \eqref{1} is integrated using a second-order
scheme given by Vanden-Eijnden and Ciccotti\cite{ciccotti} whereas the first-order equation \eqref{label04} is integrated using Euler-Maruyama Scheme \cite{EM}. These two equations collectively introduce three timescales namely the persistence time $\tau_p = 1/D_r$, inertial time scale $\tau_d = m/\gamma$ as well as meantime between collisions, the interplay of which govern the whole dynamics.
%\begin{tabular}{@{}ll@{}}
%\topple
%\textbf{Persistence time}            & $\tau_p = 1/D_r$ %\\ \midrule
%\textbf{Mean time between collisions} & $\tau_c = \pi %%\sigma / (4 v_0 \phi)$ \\ \midrule
%\textbf{Inertial time or translational memory}                
%\end{tabular}

%\subsubsection{Parameters and Initial Configuration}

We simulate a system of  $ N = 625$ iABPs in two dimensions,  initially arranged in a square lattice subjected to periodic boundary conditions with length of the simulation box $L = 250\sigma$, that corresponds to a global packing fraction $\phi = 7.85 \times 10^{-3}$. This is measured as $\phi = \frac{\pi \sigma^2 N}{4L^2}$. We refer to this as the low packing fraction scenario to distinguish it from simulations conducted at a higher packing fraction of  $\phi = 7.85 \times 10^{-2}$ to investigate the effects of higher packing fractions on the system dynamics. Here we measure time in units of $\tau = \sqrt{\frac{m \sigma^2}{k_B T}}$, energies in units of the thermal energy $k_B T$
and lengths in units of $\sigma$. The activity of iABPs are characterised by dimensionless Péclet number $\mathrm{Pe} = \frac{\sigma v_{0}}{D_{T}}$ representing the ratio of times spent in advective and diffusive motion.  Since the frictional coefficient and rotational diffusion constant determine persistence and inertial timescale respectively; we choose their values judiciously enough so that it ensures that the system does not undergo MIPS. We select $\gamma = 10^{2} \sqrt{\frac{m k_B T}{\sigma^2}}$ and $D_R = 8 \times 10^{-2} \tau^{-1}$,  resulting in $\tau_d = 10^{-2} \tau$. This choice ensures that the effect of inertia is negligible, making the system strongly overdamped\cite{,negi,mandal}.  Additionally we set $\epsilon/k_{B}T = (1 + Pe)$ to ensure a nearly
constant iABP overlap upon collisions, even at high activities\cite{negi}. We have chosen system of high activity with $Pe = 200$,$\frac{\Omega}{D_{R}} = 62.5$, and characteristic length $R_{0} = 1.5\sigma$. The simulation is performed over total $10^{7}$ realisations, while equilibrating upto $10^{6}$ steps, with timestep being $\delta t = 0.001\tau$. The trajectories of the particles are stored at an interval of  100 steps after the system goes into a steady state. We have carried out three independent simulations and then averaged over dynamical properties to improve the statistics. These trajectories are used to calculate the structural and dynamical properties of the IABP system.

\section{Results and Discussion}

\subsection{Emerging structures}

\begin{figure}
	\centering
	\subfloat[]{\includegraphics[width=0.45\linewidth]{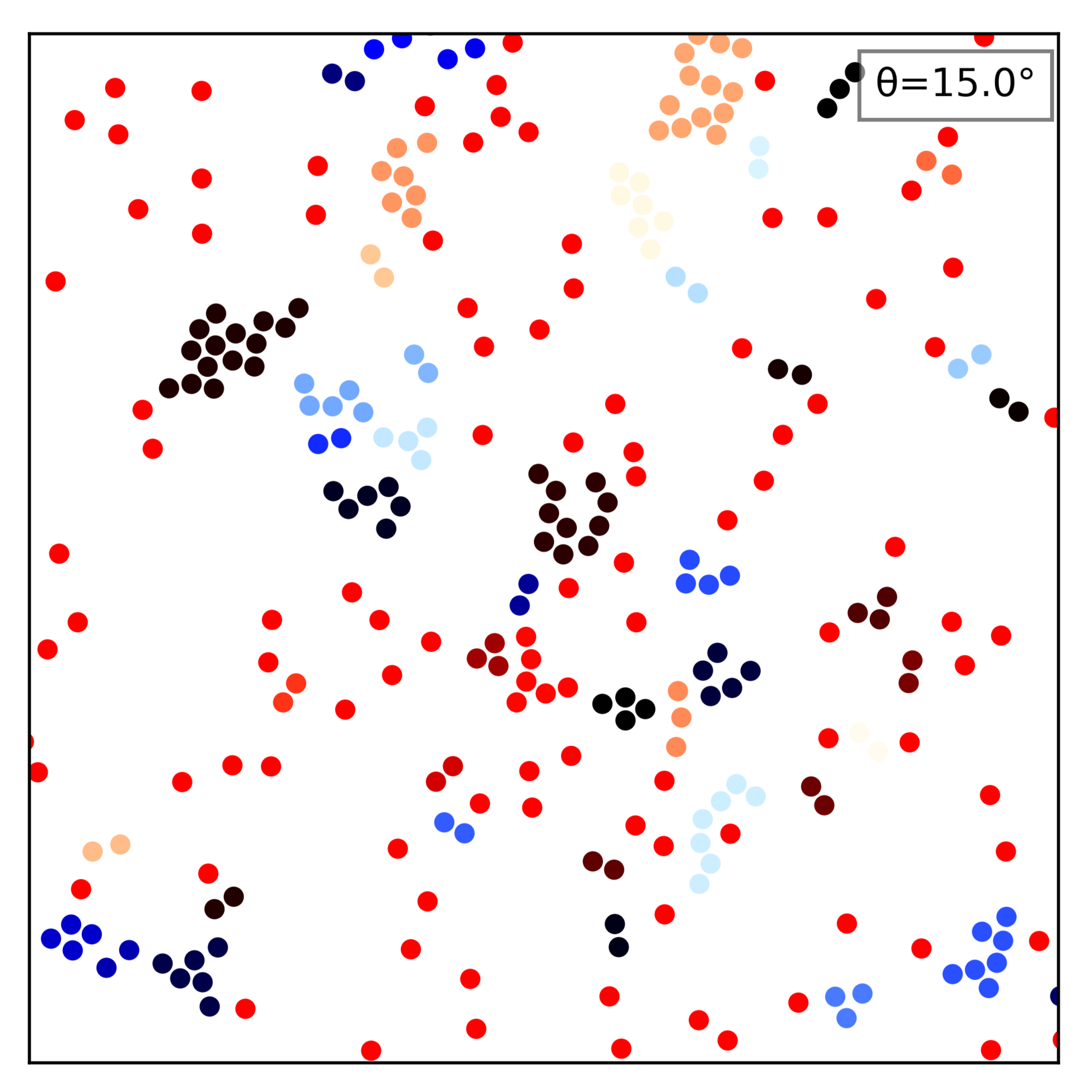}}
	\subfloat[]{\includegraphics[width=0.45\linewidth]{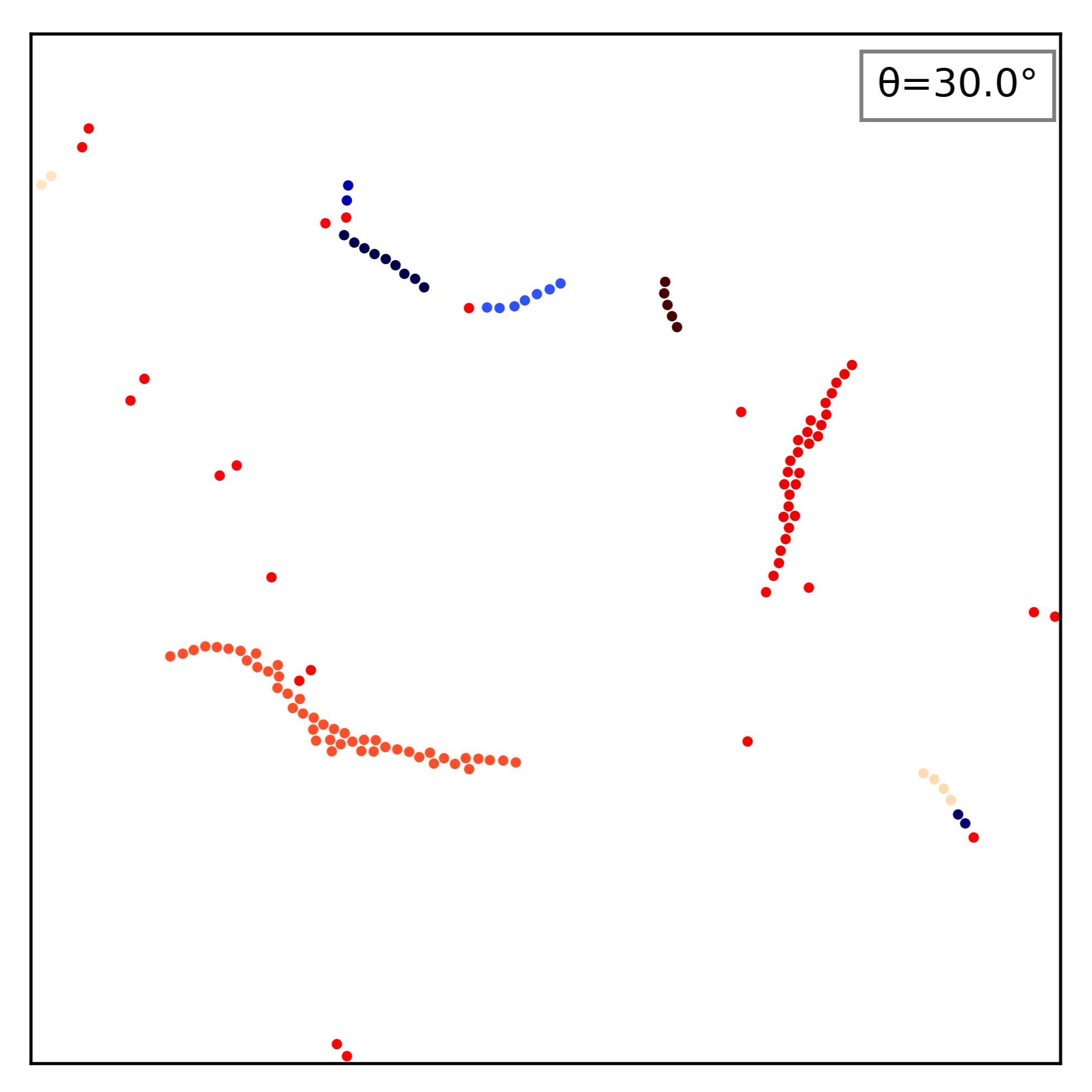}}\\
	\subfloat[]{\includegraphics[width=0.45\linewidth]{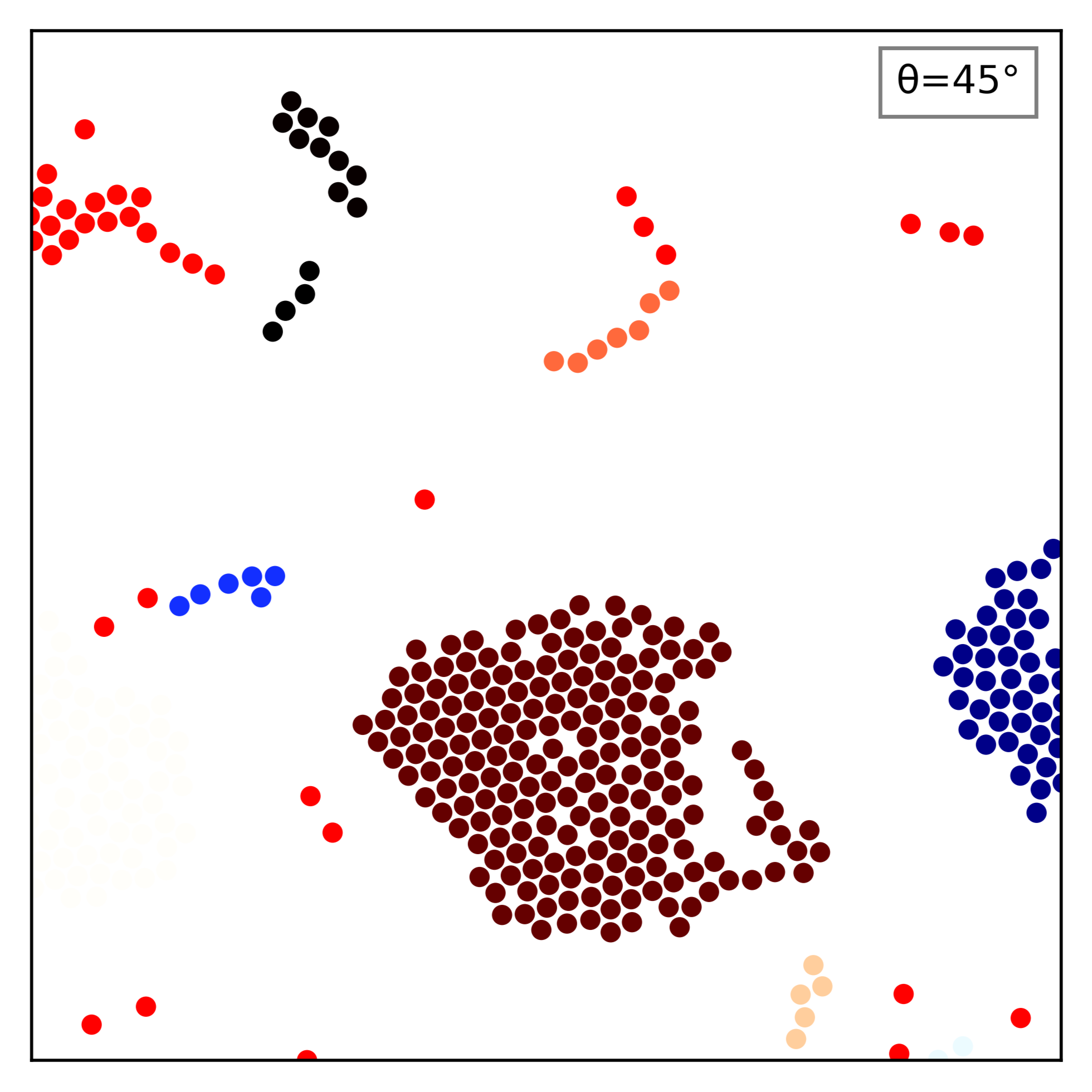}}
	\subfloat[]{\includegraphics[width=0.45\linewidth]{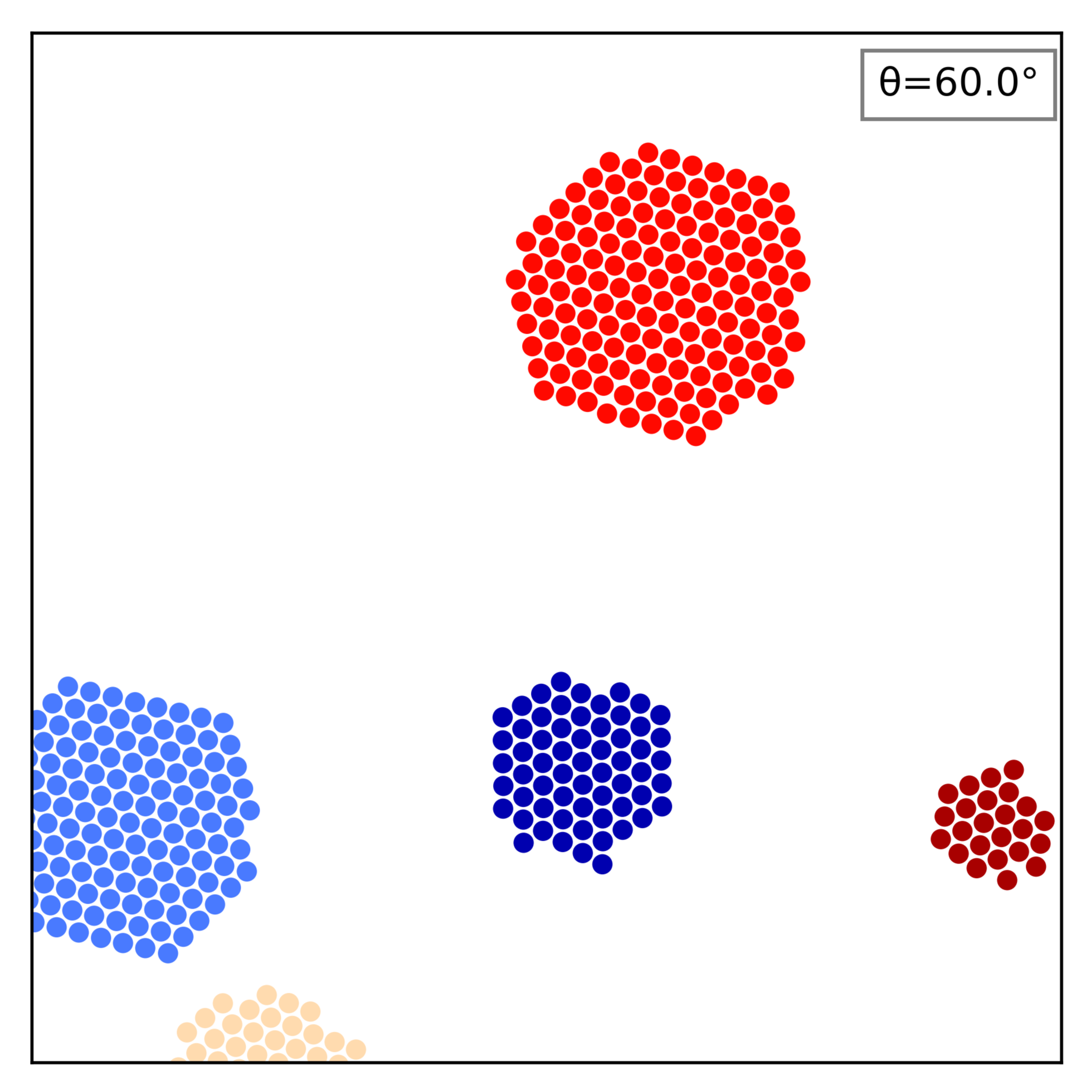}}
	\caption{Different emergent structures at various vision angles at lower packing fraction: (a) $\theta = 15^\circ$, (b) $\theta = 30^\circ$, (c) $\theta = 45^\circ$, and (d) $\theta = 60^\circ$.}
	\label{fig:struct1}
\end{figure}

 As stated in the earlier section, we have carried out the Langevin dynamics simulations of iABPs at two different packing fractions. For each packing fraction, the simulations were done for a range of vision angles, $\theta$  from $\pi$/18 to $\pi$/2. For the low packing fraction, we have observed different emerging structures as the vision angle is varied. Snapshots of various structures at representative vision angles are depicted in Figure \ref{fig:struct1}. For very low vision angles, the system remains as a dilute fluid, where the particles exhibit a random and disordered arrangement with minimal clustering or alignment.  However with the increase in the vision angle $\theta$, the particles gets aligned by adaptive vision induced torque giving rise to various structures.  Worm-like structures, i.e., an elongated clusters of particles that exhibit a higher degree of alignment and directional motion starts emerging from $\theta$ = $\pi$/9 and becoming more distinctive near $\theta$ = $\pi/$6. With further increase in $\theta$, the worms self-collide or collide with other worms and start forming larger aggregates. This will first lead to a phase where worms and larger aggregates coexists, for example, for $\theta$ = $\pi$/4. Further increase in vision angles leads to larger aggregates or clusters which are hexagonal-close packed structures.  At higher packing fractions, say $\phi$ = 0.0785, we observe similar phases. However, the structures starts emerging at even smaller vision angles. Our results are in good agreement with the structures reported by Negi {\it et al.}\cite{negi} 
 
\subsection{Diffusion}

\begin{figure}
    \centering
    \includegraphics[width=0.90\linewidth]{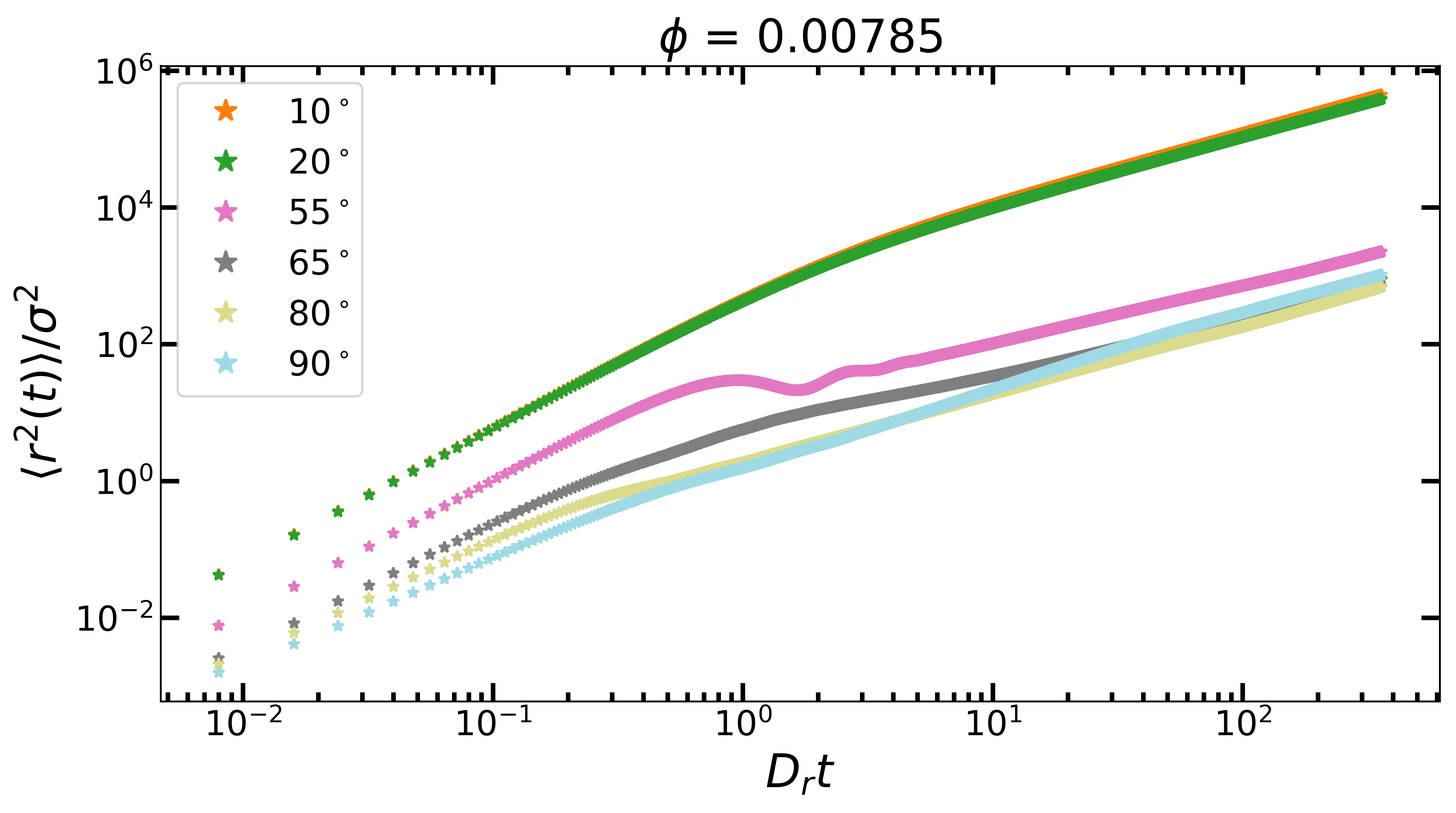}
    \caption{ Mean Square Displacement for various vision angles at packing fraction $\phi$ = 0.00785 .}
\label{fig:msd}
\end{figure}

\begin{figure}
    \centering
    \includegraphics[width=0.90\linewidth]{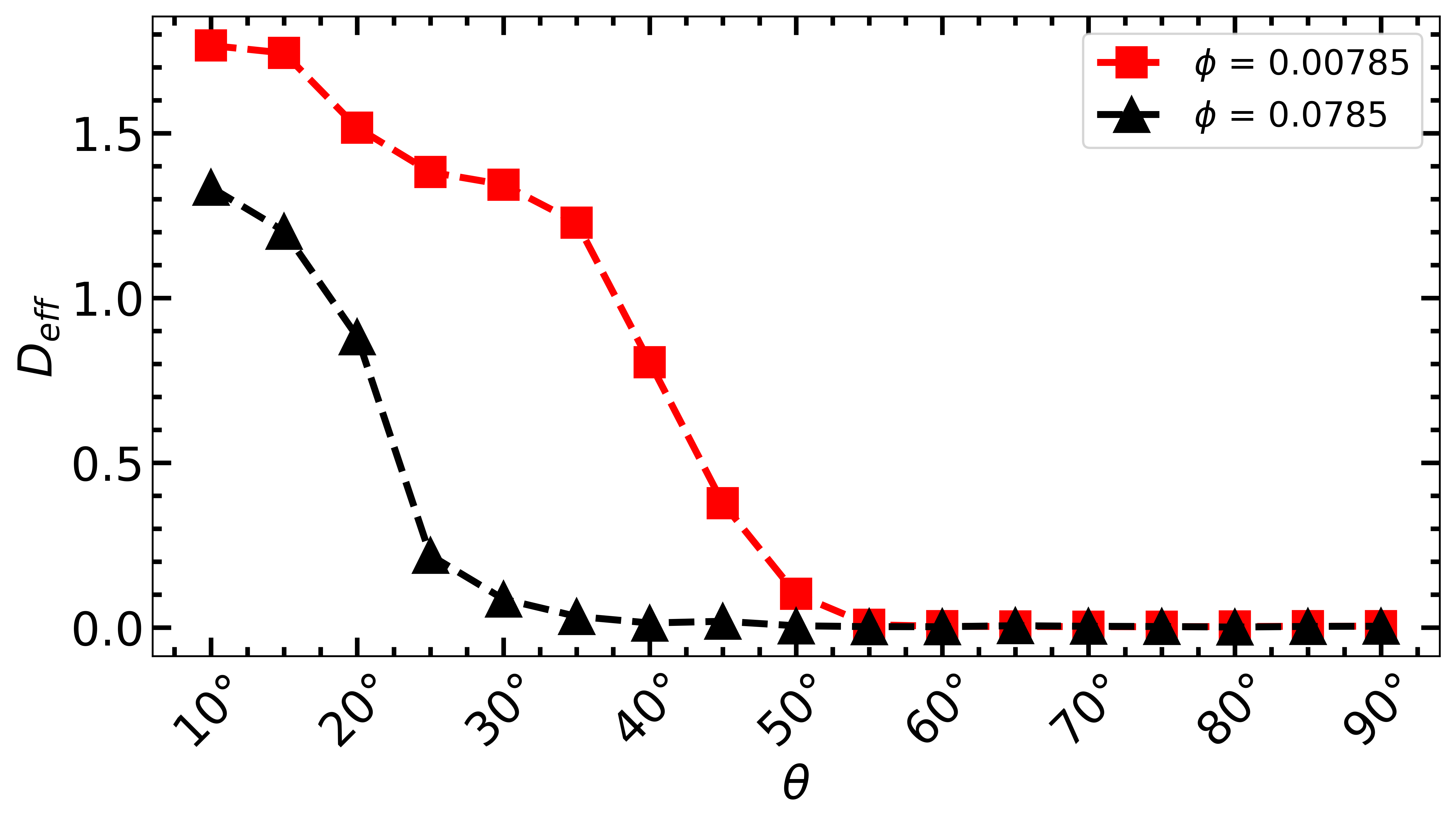}
    \caption{Effective diffusion coefficient $D_{eff}$ for various vision angles (a) $\phi$ = 0.00785 and  (b) $\phi$ = 0.0785}
\label{fig:diff}
\end{figure}

 To characterise the dynamics of iABPs, we have calculated the particle's mean squared displacement. This is depicted in Figure \ref{fig:msd}. At large times, the mean squared displacement approaches a linear behaviour in time and the dynamics becomes diffusive for all the vision angles.  It must be noted that for the intermediate vision angles, mean squared displacement shows an oscillatory  behaviour at intermediate time scales before going in to the linear regime. We have calculated the effective  longtime diffusion coefficient of iABPs from their mean squared displacement. This is plotted in Figure \ref{fig:diff} against the vision angles for the two packing fractions we have investigated.  As evident from the figure, it is clear that there are two distinct regimes in the dynamics here. For very low vision angles, the diffusion coefficient is higher. Then towards the mid range of vision angles, there is a sharp decrease in the diffusion coefficient.  As discussed above,  the structure of iABPs change from dilute fluid to aggregates around the mid values of vision angles. The particles in these aggregates move together, so we can consider them as single bigger particle of larger size and mass moving.  This correlated motion will be obviously slower than the individual iABPs in the dilute fluid phase. This results in a lower diffusivity, which is reflected in Figure \ref{fig:diff}.
 
\subsection{Rotation of Clusters}

\begin{figure}
	\centering
	\subfloat[]{\includegraphics[width=0.45\linewidth]{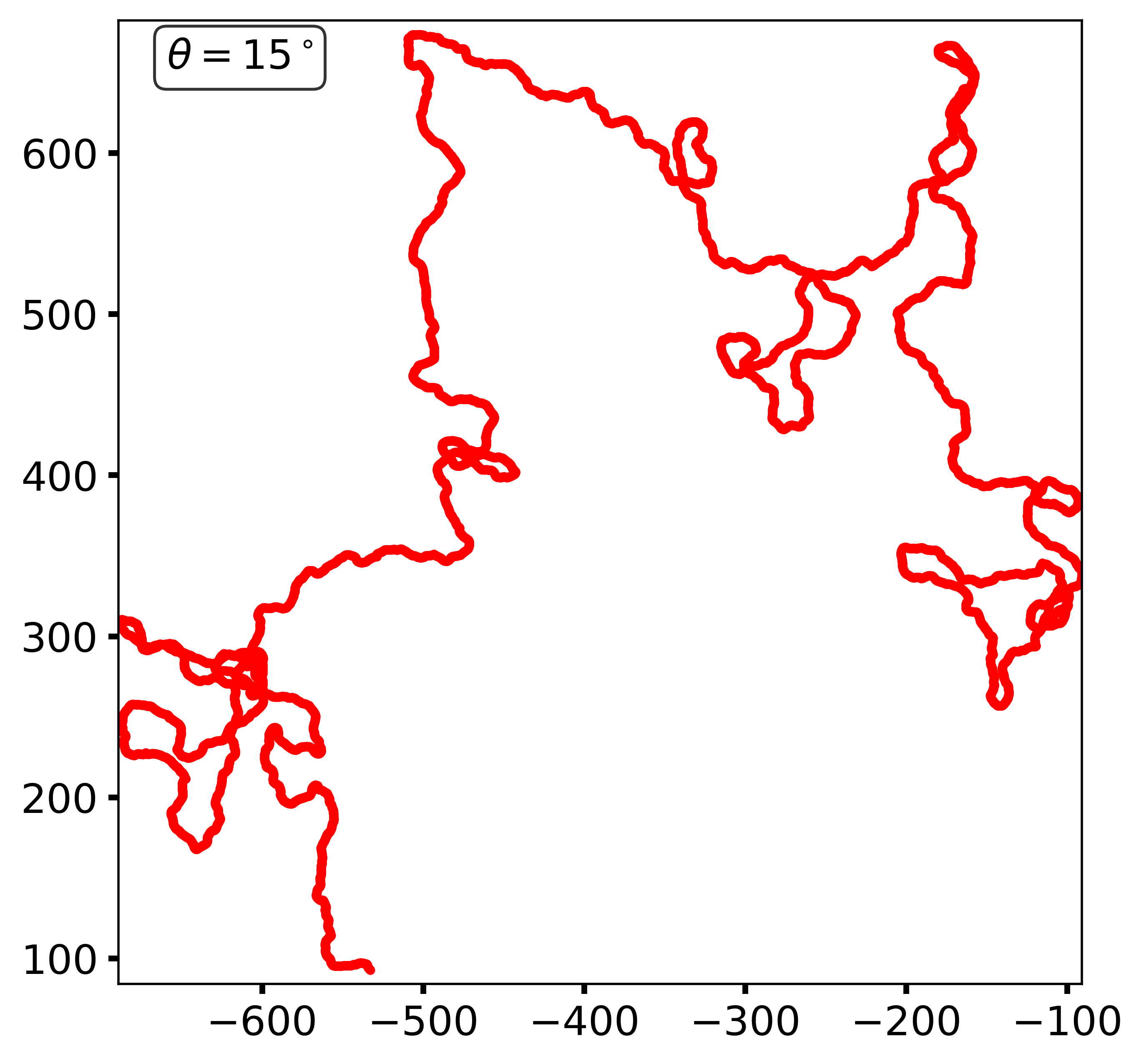}}
	\subfloat[]{\includegraphics[width=0.45\linewidth]{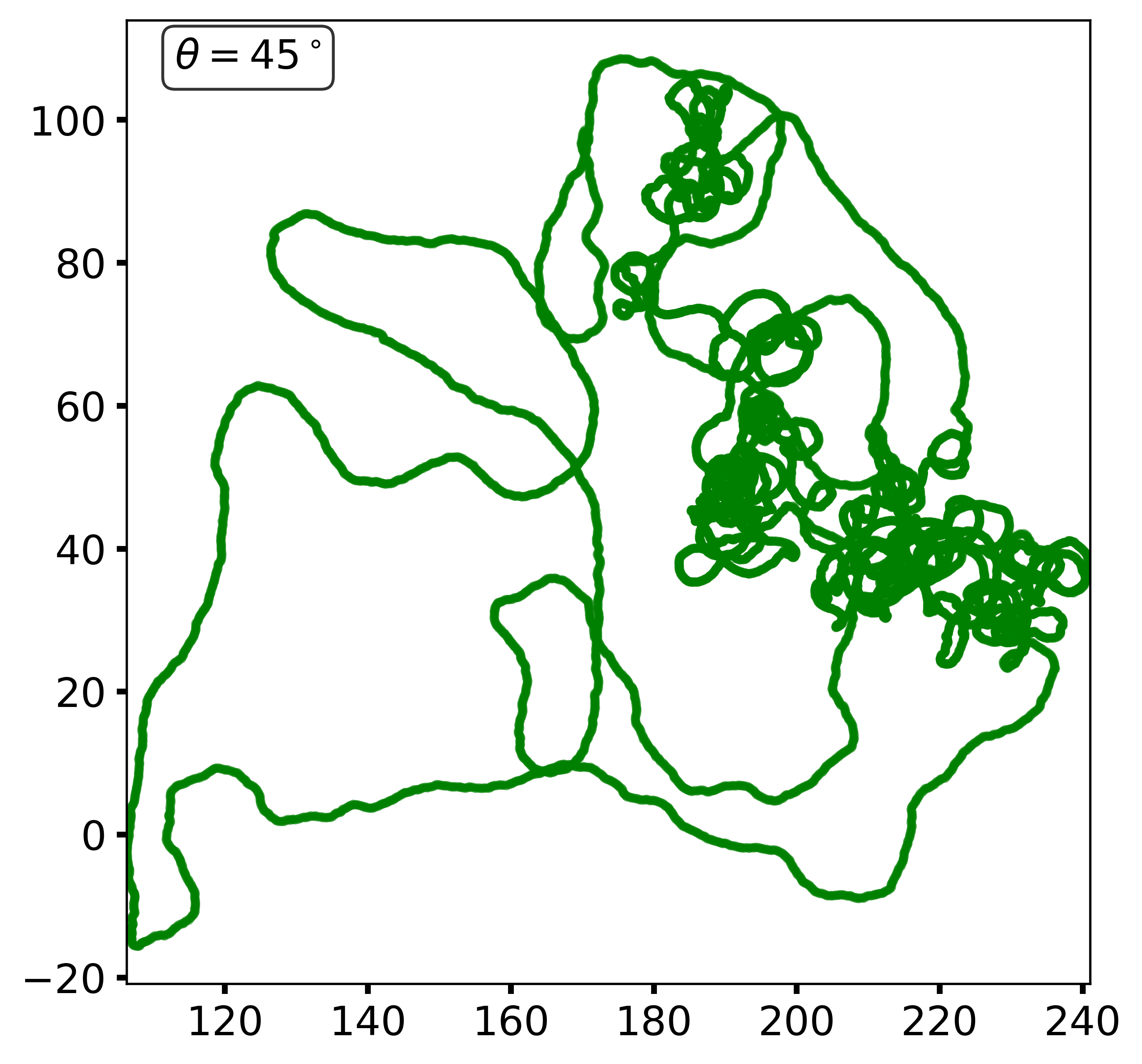}}\\
	\subfloat[]{\includegraphics[width=0.45\linewidth]{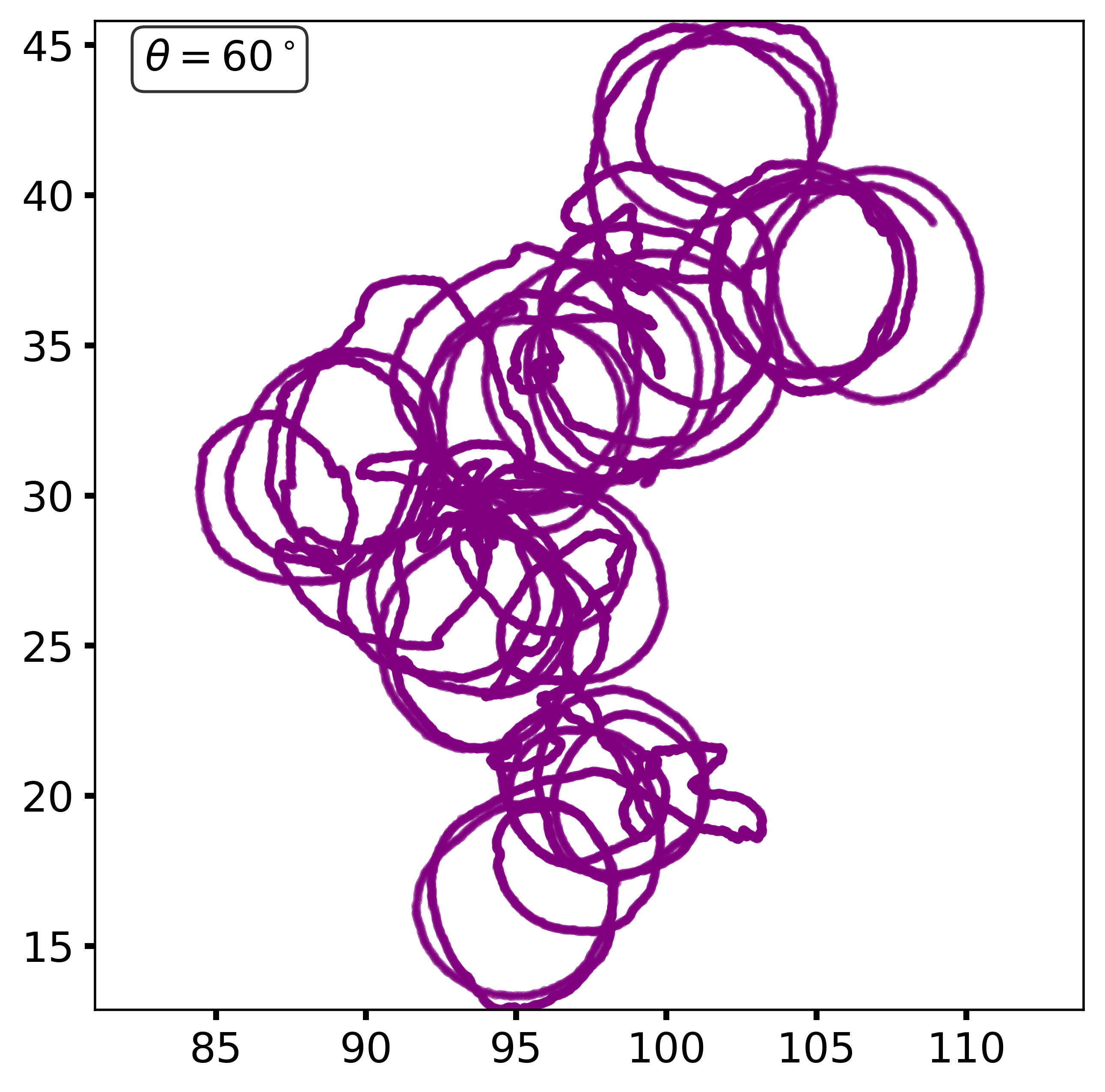}}
	\subfloat[]{\includegraphics[width=0.45\linewidth]{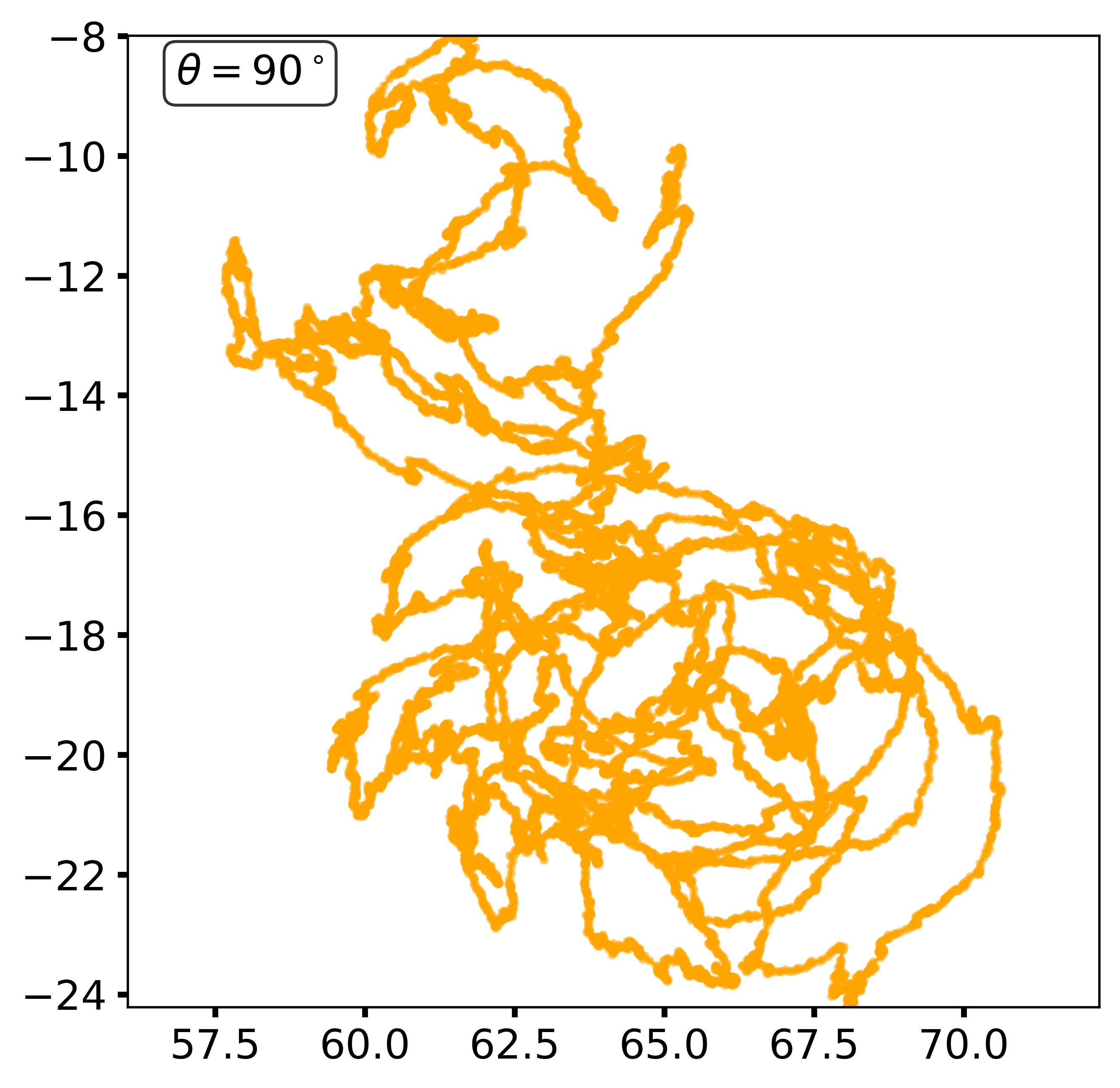}}
	\caption{Representative single-particle trajectories at various vision angles at lower packing fraction: (a) $\theta = 15^\circ$, (b) $\theta = 45^\circ$, (c) $\theta = 60^\circ$, and (d) $\theta = 90^\circ$.}
	\label{fig:trajectory}
\end{figure}

 Despite the diffusive behaviour at longer times, the mean squared displacement, especially at intermediate vision angles shows an oscillatory behaviour for timescales, $D_rt \sim$ 1. In order to further investigate this, we have scrutinised the individual particle trajectories. Figure  \ref{fig:trajectory} depicts the representative trajectory types for four different vision angles, where we find different type of emerging structures.  Figure \ref{fig:trajectory}(a) represents the single particle trajectory at vision angle $15^\circ$. At this vision angle, the system is a dilute phase and the trajectory reveals a non random path, typical of active particles.  In Figure \ref{fig:trajectory}(b), the trajectory shown is at a vision angle where the worm phase and hexagonal cluster coexists. This trajectory shows an interesting behaviour. It has  long persistent motion part interlaced by curly bounded part. Since the worm structure and hexagonal cluster are very dynamic in time, particles will be getting attached to and detached from these structures dynamically. The curly bounded part of the trajectory correspond to the time when the particle is part of a hexagonal cluster when the movement of cluster is slow and correlated, while the long persistent motion part is when the particle get detached from the cluster or when it is part of the worm structure.  This slow dynamics of the hexagonal clusters in this co-existing phase decreases the effective diffusivity of the system. This is the region in Figure \ref{fig:diff}, where we observe a sharp decline in the diffusivity.  The third trajectory shown in Figure \ref{fig:trajectory}(c) is representative trajectory of a particle for the vision angle $60^\circ$.  At this vision angle, the trajectory is curly, almost circular and these circular orbits are moving translationally.  As discussed above,  the  active particles form hexagonal clusters at this vision angle.  The nature of the trajectory  indicates that these clusters are rotating as well as translating.  We have scrutinised the time evolution of configurations and found that the hexagonal clusters are indeed rotating(see the movie M1.gif at \cite{suppinfo}).  Such rotating clusters are earlier observed in active systems\cite{yang,huang,cantisan,kokot}.  Generally collective rotations can be achieved by  an external field such as magnetic field\cite{grzybowski,yan} or  optical tweezers\cite{grier, moffitt}.  Biological organisms such as dancing algae\cite{drescher} or sperm cells\cite{riedel} can form rotating clusters naturally.  In most of these situations,  rotating clusters are formed by anisotropic active particles\cite{kummel}. In our model system, there is no anisotropy in the particle shape as all the particles are modelled as circular discs. Anisotropy in the rotational motion of each particles is brought in to the system by the visual perception.  However, it must be noted that the rotating clusters are observed for a certain range of vision angles which suggest that there is certain threshold of torque above which the clusters start rotating. Having a smaller or very large vision angle randomises the rotational motion of the active particle which brings down the total torque of the cluster below the threshold, resulting in no significant rotational motion. The trajectories at even higher vision angles(shown in Figure \ref{fig:trajectory}(d) for $90^\circ$)  again comes back to random diffusive trajectories. This suggests that the clusters formed at  higher visional angles do not rotate and undergoes only translational motion(see the movie M2.gif at \cite{suppinfo}). So the analysis of trajectories as well as their visualisation indicates the hexagonal clusters formed at intermediate range of vision angles rotates, while those at higher vision angles undergoes only translational motion.
 
\begin{figure}
	
	\centering
	\subfloat[]{ \includegraphics[width=0.9\linewidth]{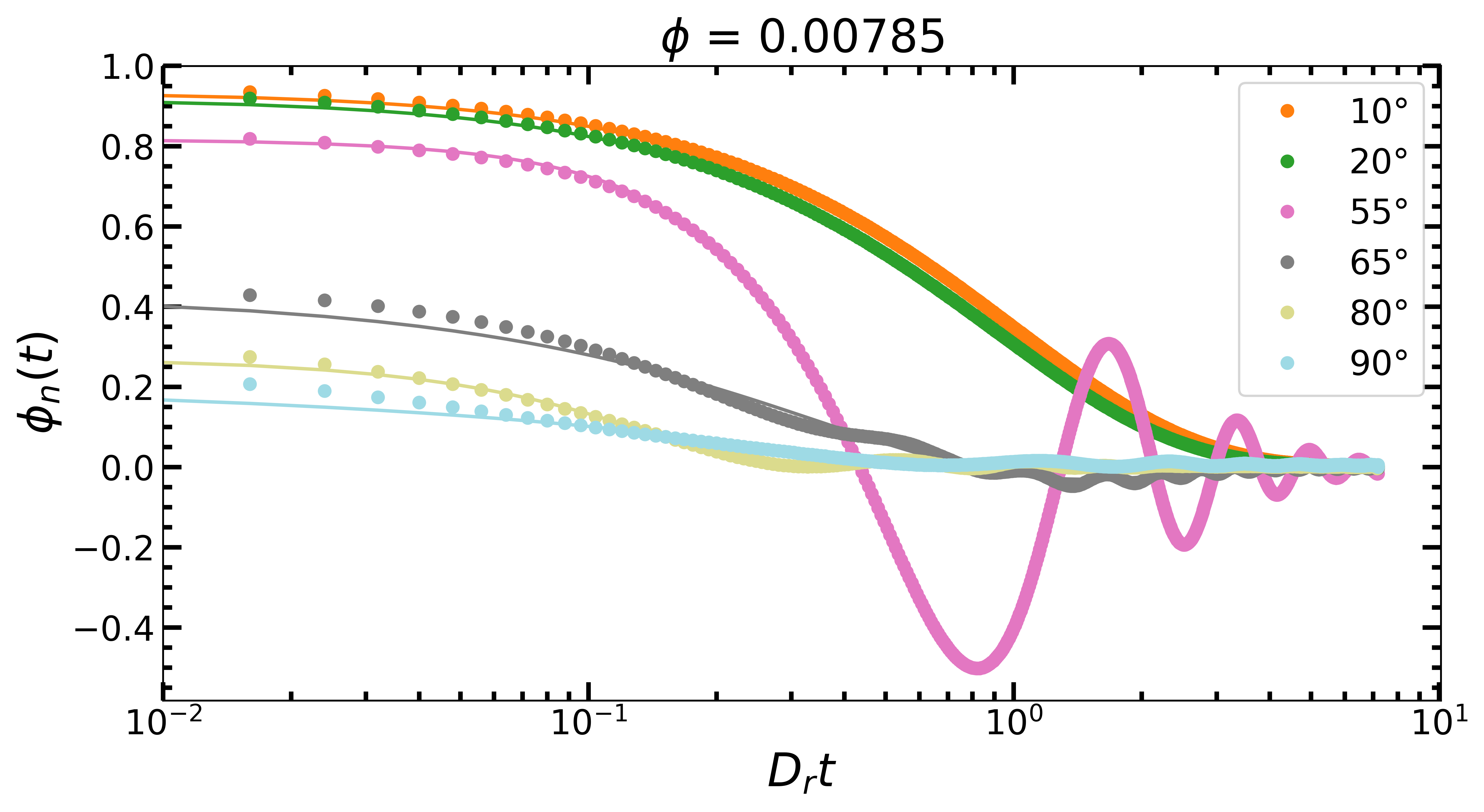}}\\
	\subfloat[]{ \includegraphics[width=0.9\linewidth]{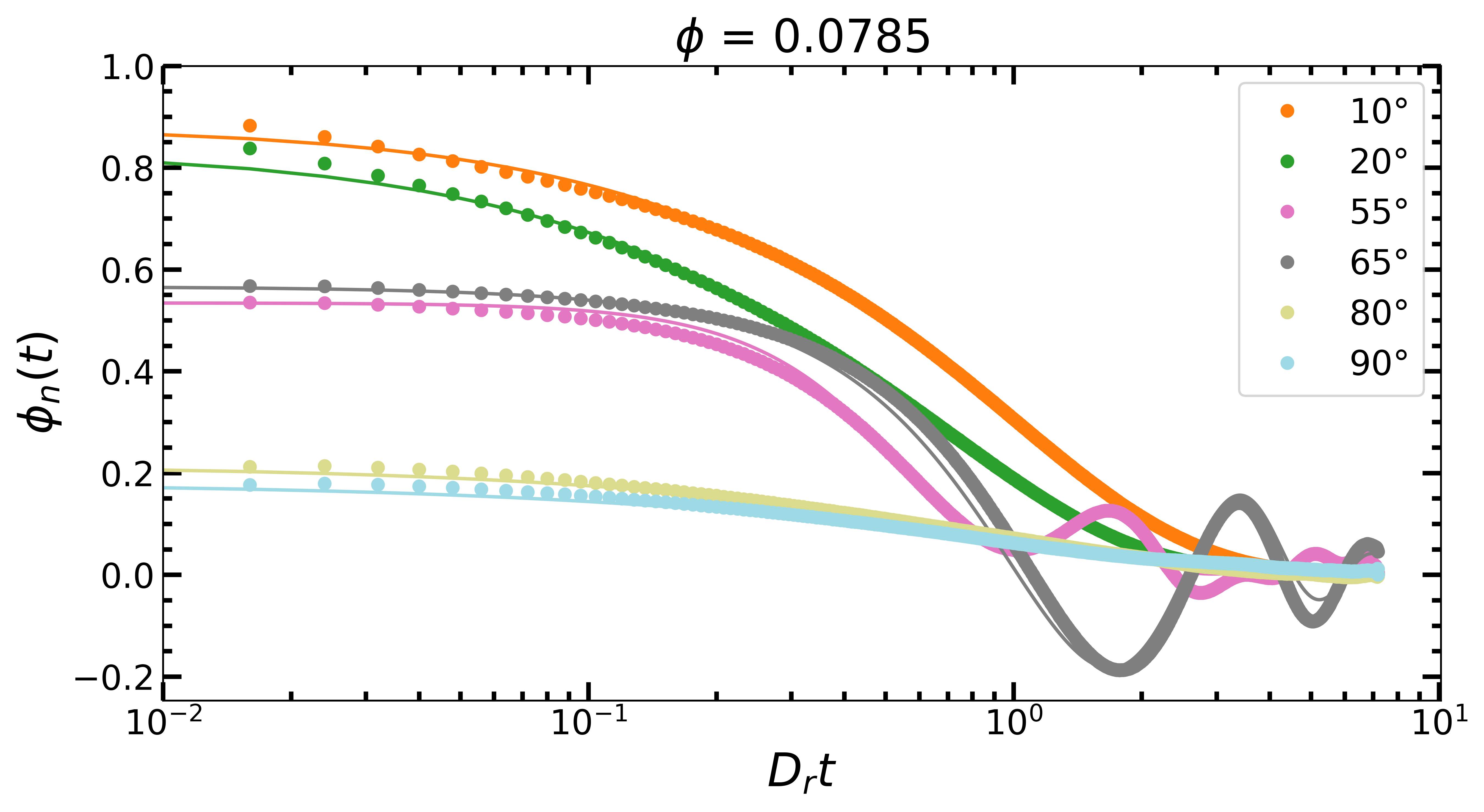}}
	\caption{Directed Autocorrelation for various vision angles at both packing fractions (a) $\phi$ = 0.00785 and  (b) $\phi$ = 0.0785 .}
	\label{fig:autocorrelation}
\end{figure} 
 
To further characterise these rotating clusters, we have calculated the directional autocorrelation function, defined as\cite{ vaccari}
\begin{equation}
\phi_n(t) = \left\langle \mathbf{n}(t' + t) \cdot \mathbf{n}(t') \right\rangle_{t'}
\end{equation}

Here $\mathbf{n}(t')$ is the unit vector along the direction of particle motion at time $t'$.  This correlation function allows us to determine the shape of the trajectories individual particles are following.  For a passive particle, undergoing diffusive motion, the direction autocorrelation function decays exponentially, i. e, $\phi_n(t) \sim e^{-t/\tau_c}$, where $\tau_c$ is the correlation time.  However in active particles, the trajectories take more time to randomise and the directional autocorrelation function can take nontrivial forms. In Figure \ref{fig:autocorrelation}, we have plotted the  directed autocorrelation function for some representative vision angles at the two packing fractions we have investigated.  For both the packing fractions, at lower and higher vision angles, we have observed that the directional autocorrelation function is exponentially decaying.  For some vision angles we found that the orientational autocorrelation function decays as a stretched exponential way  rather than as an exponential.  Also, for the mid range of vision angles($45^\circ$ - $70^\circ$), the directional autocorrelation function exhibit pronounced oscillations.  These oscillations suggest that the individual particles follow  curly or circular  trajectories. Thus, the autocorrelation function is a combination of exponential decay and a periodic oscillatory function. So we fitted all the directional oscillatory function with a function of the following form
\begin{equation}
\phi_n(t) = A \Big(cos (2\pi\omega t) + c\Big)e^{-(t/\tau)^{\beta}}
\end{equation}
\noindent which can take into account of all these behaviours.  The fit is shown as the continuous lines in the Figure \ref{fig:autocorrelation}, whereas the symbols are from the simulations.  Since there are five different parameters to be fitted in this equation, listing out the dependence of these parameters on the vision angles and forming conclusions about the rotating clusters is rather a tedious and difficult task.  In order to simplify this, we have defined a metric combining multiple parameters as

\begin{equation}
S = \frac{A}{|c|+1}\omega \tau
\end{equation}

\begin{figure}
 
    \centering
   \includegraphics[width=0.9\linewidth]{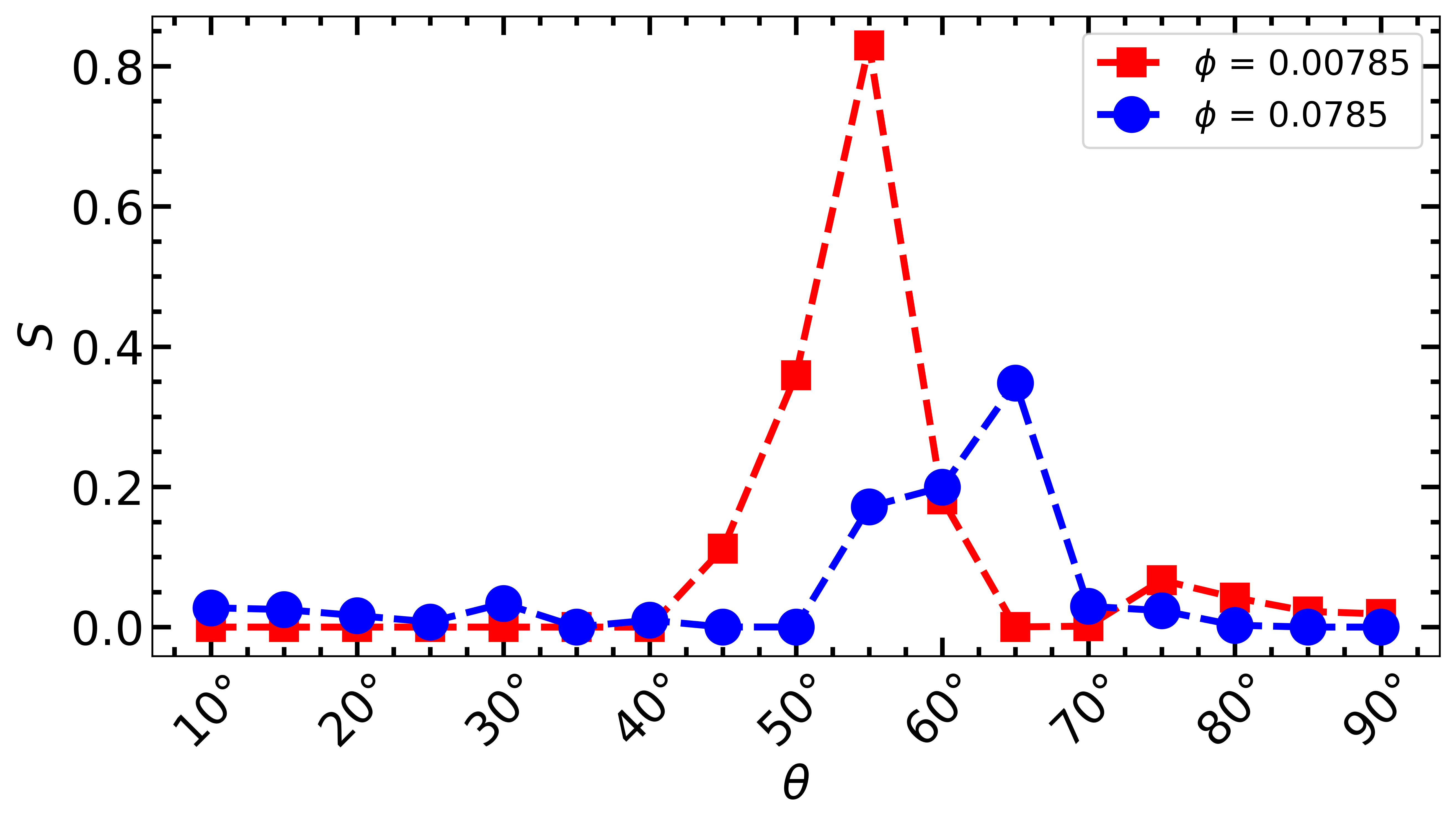}
        \caption{Metric at various vision angles at both packing fractions (a) $\phi$ = 0.00785, (b) $\phi$ = 0.0785 .}
\label{fig:metric}
\end{figure}

\noindent Unlike simple frequency based measures, the metric $S$, considers both strength and persistence of oscillations, making it more representative of directional autocorrelation function. Here $A$ is the amplitude of oscillations and its larger values allow the oscillations to persist for a longer time; hence $S$ is directly proportional to $A$.  The oscillatory component in the fit is given by $cos(2\pi \omega t)+c$. If $c$ is large, the cosine term gets overshadowed by $c$ suppressing the oscillations in the fit function. So dividing by the absolute value of $|c|$ corrects this suppression. $\tau$ is the decay rate in the exponential which determines the persistence of these oscillations. If $\tau$ is very small, the oscillations die out very quickly and oscillations persists for larger values of $\tau$. This is reflected in the metric $S$ as a linear dependence on $\tau$.  We have plotted this metric $S$ against the vision angle for both the packing fractions we simulated in Figure \ref{fig:metric}.  As evident from the figure, $S$ shows a peak in the mid ranges of vision angles  for both packing fractions. At both ends of the vision angles, $S$ is very close to zero, indicating the no oscillations in the orientational correlation function and the non-existence of rotating clusters.  However  $S$ shows a peak in the mid ranges of vision angles  for both packing fractions, suggesting the emergence of rotating clusters and the subsequent oscillatory behaviour in the orientational correlation function.  The range at which the rotating clusters exists get shifted slightly towards the larger values as the packing faction increases. It should be noted that we have not included the stretched exponential exponent $\beta$ in the definition of $S$, since $\beta$ affects only the decay shape and does not affect the oscillatory strength directly.  

\begin{figure}
    \centering
    \subfloat[]{ \includegraphics[width=0.9\linewidth]{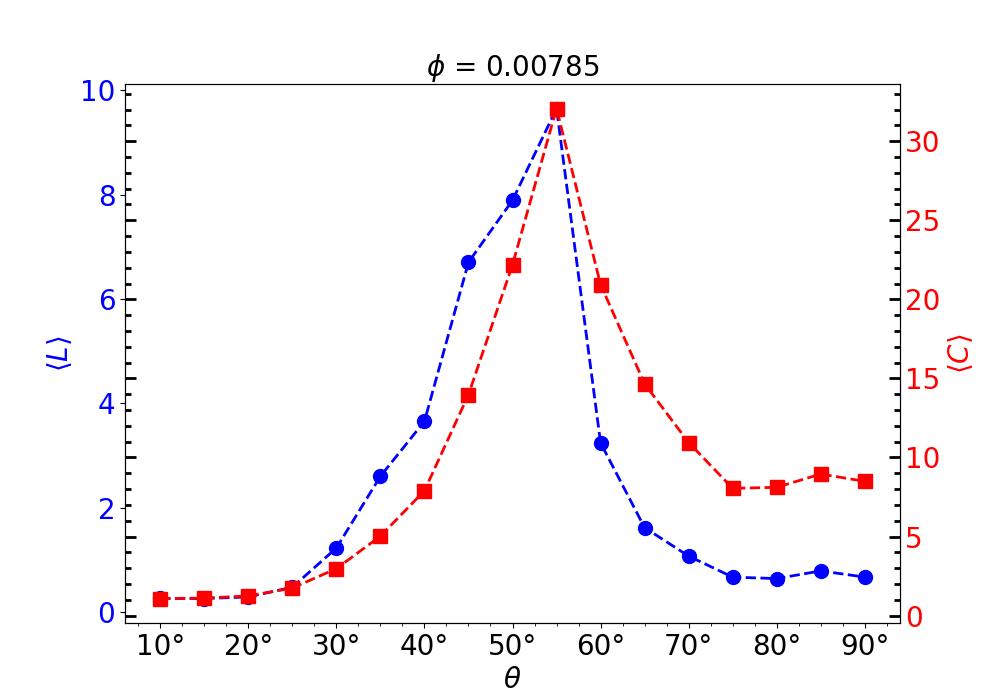}}\\
    \subfloat[]{ \includegraphics[width=0.9\linewidth]{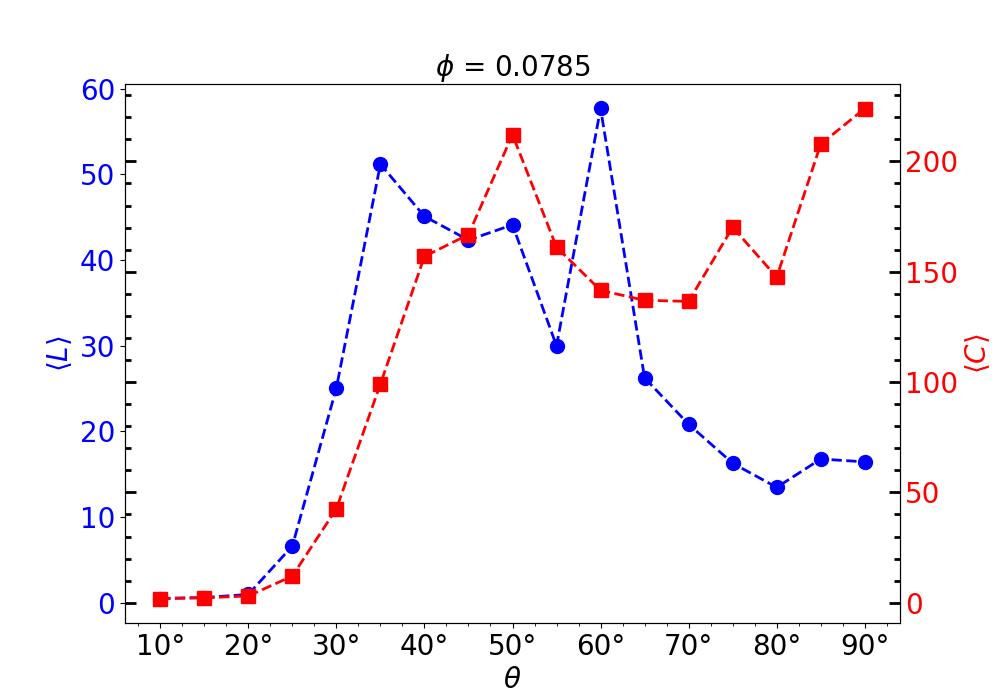}}
    \caption{Mean cluster size and average angular momentum  for various vision angles at both packing fractions (a) $\phi$ = 0.00785 and  (b) $\phi$ = 0.0785 .}
\label{fig:cluster}
\end{figure}

From the discussions above, it is clear that the dynamics of iABPs with a mid range of vision angles is quite different from other range of vision angles.  At this range of vision angles,  the active particles form hexagonal aggregates and the trajectories of the individual particles are curly or circular indicating the presence of rotating clusters. However, the aggregates formed at larger vision angles do not show any rotation.  In order to explain this observation further, we have calculated the average size of aggregates, $<C>$ as well the angular momentum, $<L>$ these aggregates for various vision angles. These are plotted in Figure \ref{fig:cluster} for both the packing fractions we have considered. It is clear from the figure that  average angular momentum is very low at both ends of the vision angle and shows a maximum at the intermediate range of vision angles.  This again validates the existence of rotating clusters at these vision angles.  For the low packing fraction, the largest clusters are formed at the mid range of vision angles.  and the peaks in the average angular momentum and  average cluster size are coinciding.  However, for the higher packing fraction,  after the peak at the mid range of vision angles the average cluster size decreases initially and increase thereafter.  This suggests that there is an optimal size of aggregates for  the rotational motion. This essentially tells us that largest clusters are formed at mid ranges of vision angles and these clusters are rotating.   

\subsection{Persistent motion}

 \begin{figure}
    \centering
    \subfloat[]{ \includegraphics[width=0.9\linewidth]{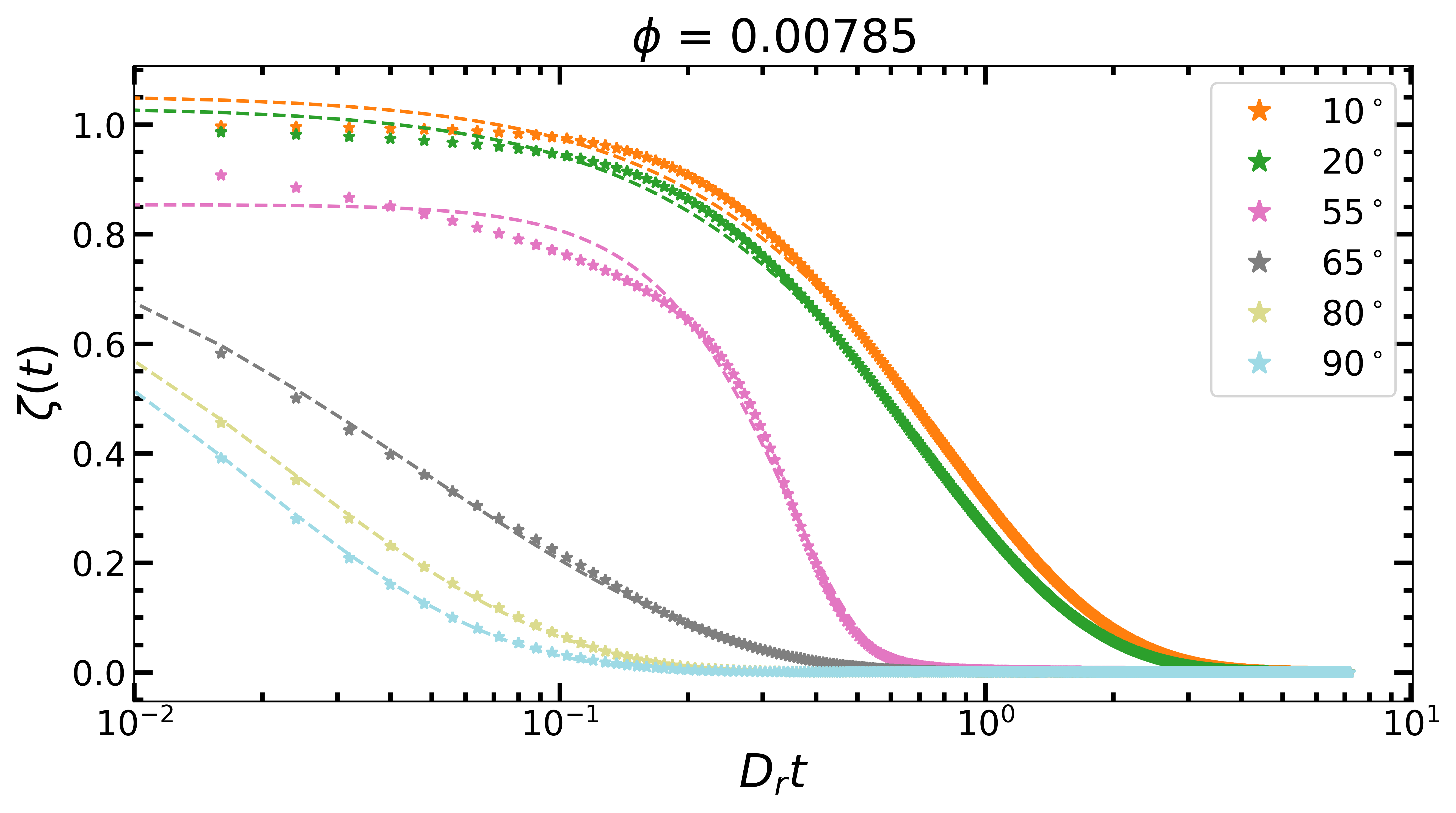}}\\
    \subfloat[]{ \includegraphics[width=0.9\linewidth]{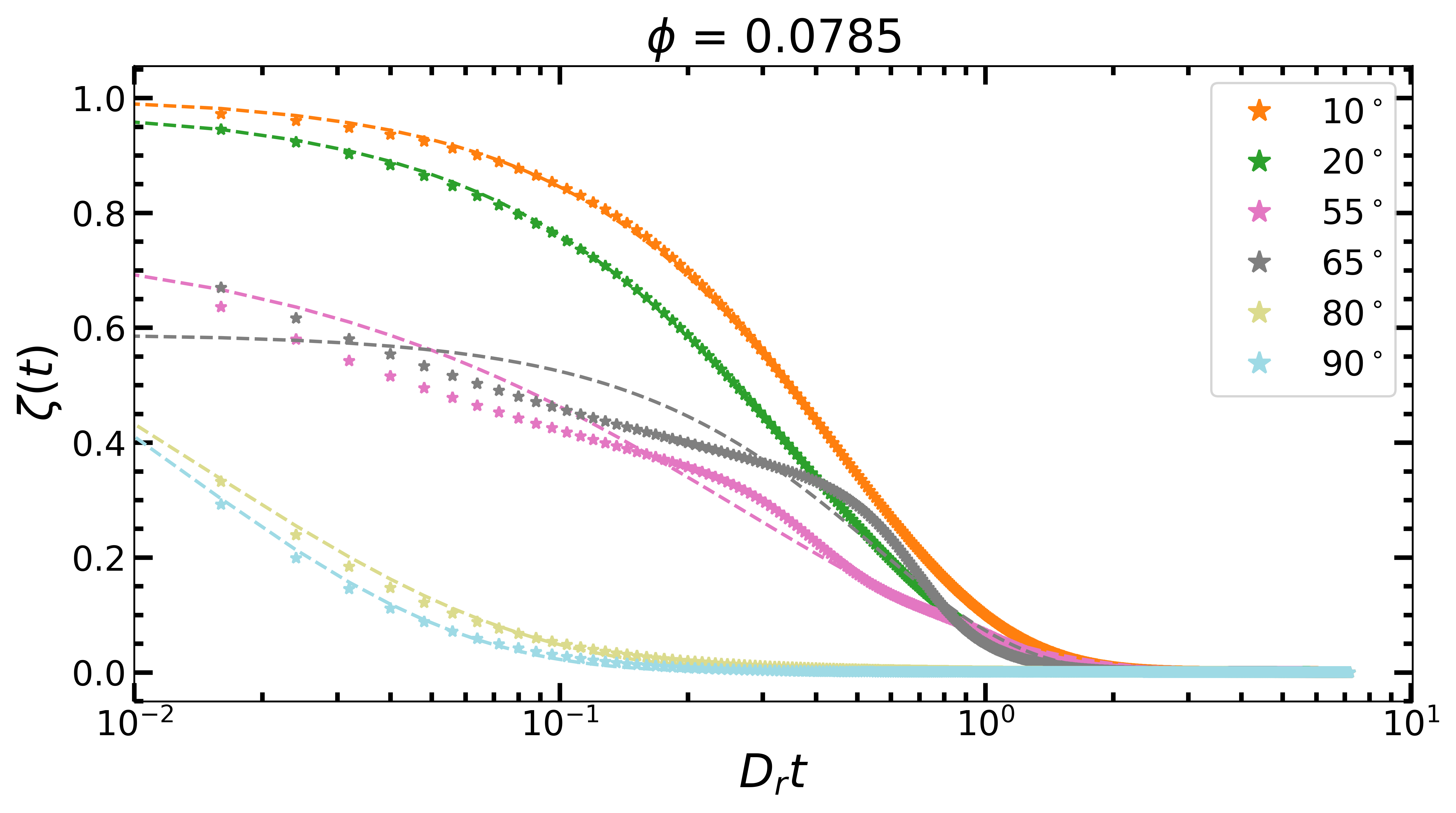}}
    \caption{Persistence probabilities at various vision angles for both packing fractions (a) $\phi$ = 0.00785 and  (b) $\phi$ = 0.0785 .}
\label{fig:persistence}
\end{figure}

\begin{figure}
    \centering
    \subfloat[]{ \includegraphics[width=0.9\linewidth]{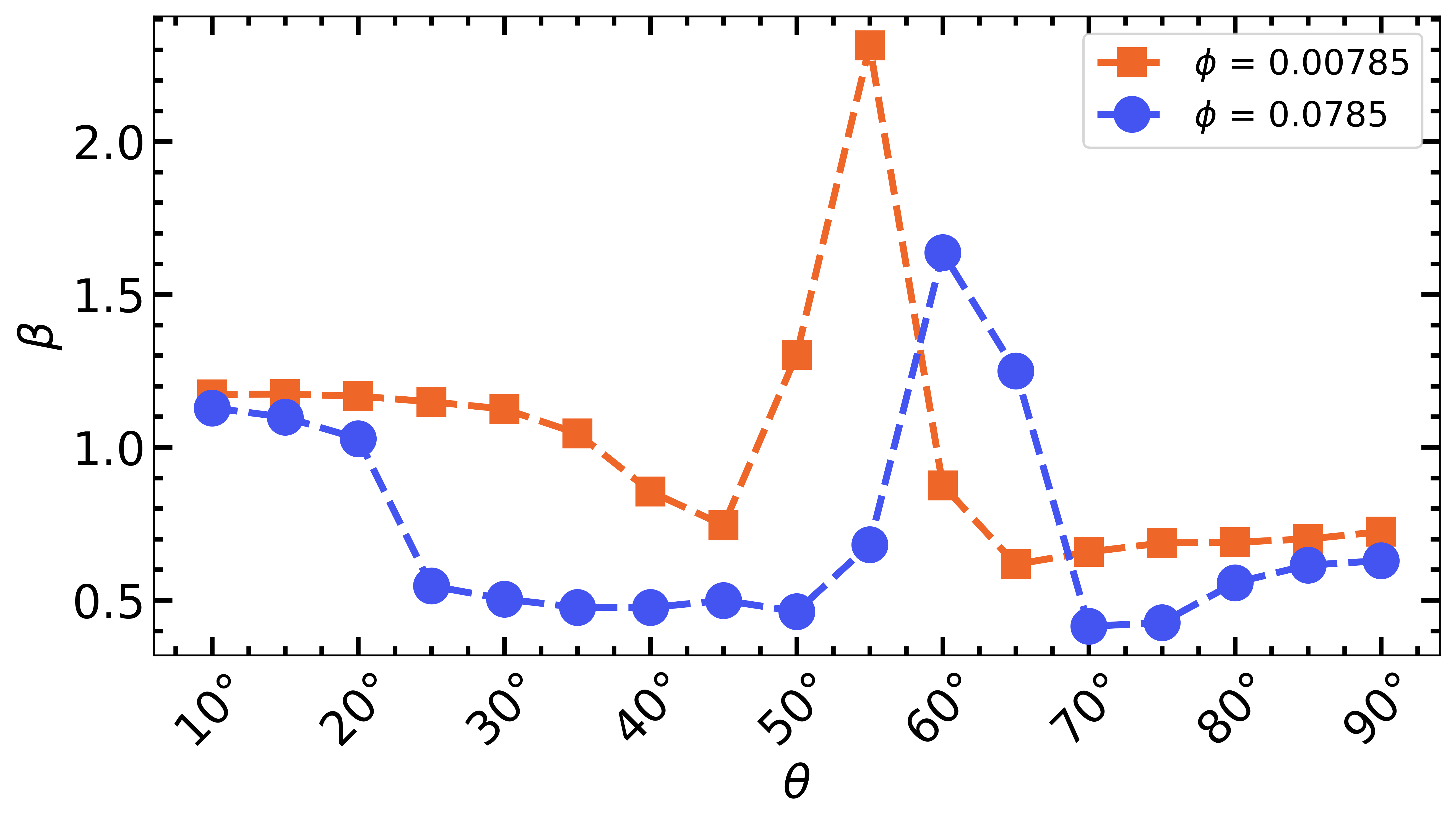}}\\
    \subfloat[]{ \includegraphics[width=0.9\linewidth]{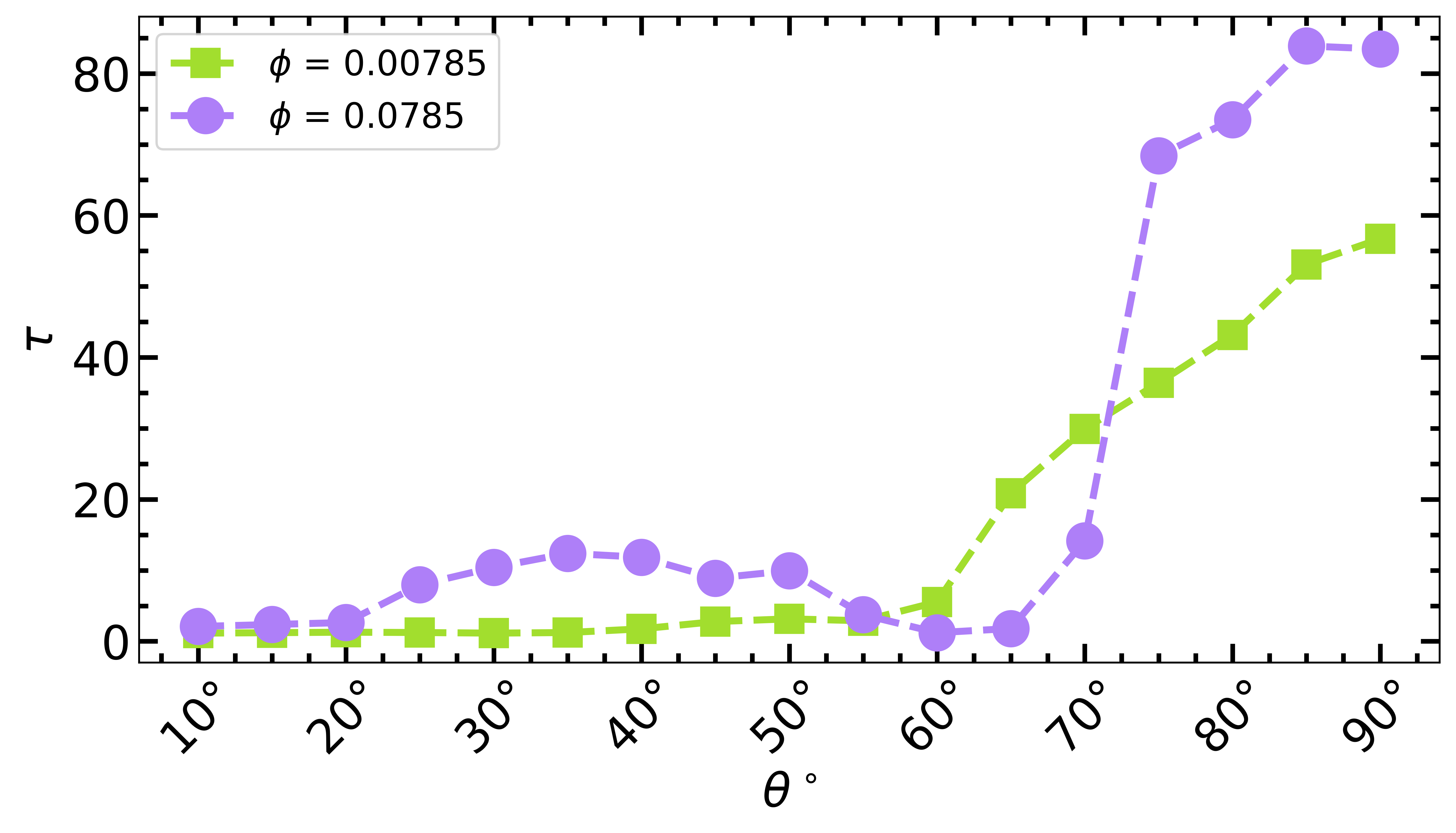}}
    \caption{Fitting parameters (a) $\beta$ and (b) $\tau$ at various vision angles for a) $\phi$ = 0.00785 and $\phi$ = 0.0785 .}
\label{fig:tau_beta}
\end{figure}

   As discussed in the previous sections, various structures emerge as the visual perception of the particles widens. This implies that the particles move cooperatively at larger vision angles.  Also, for intermediate vision angles the cluster undergoes a rotational motion as a whole. So it would be interesting to see how directed the individual motion of particles is and whether the visual perception plays a role in it.  In order to quantify this, we have defined a variable $\zeta$, which is the cosine of the angle between two displacement vectors on the trajectory of same particle. 
\begin{equation}
\zeta(t) =  {\mathbf n}(t').{\mathbf n}(t'+t)
\end{equation}
\noindent where ${\mathbf n}(t')$ is the unit vector in the direction of displacement of the particle at time $t'$. At $t$=0, the particle motion is perfectly persistent, the angle between the subsequent displacement vectors will be zero and the value of the stochastic variable $\zeta$ will be 1.  As  $\Delta t$ increases, the motion becomes less directional and the value of $\zeta$ starts decreasing from one and when the motion reverses, its value will cross to the negative side. We can define the persistence probability of this stochastic variable $\zeta$ as the probability at which the value of $\zeta$ remains positive\cite{majumdar1}. Thus persistence probability of $\zeta$ provides us a quantitative measure of the directional motion of iABPs. For a wide class of non-equilibrium systems, persistence probability has been obtained both analytically and numerically. This includes  a large class of stochastic processes which are  both Markovian as well as non-Markovian in nature, ranging from critical dynamics\cite{majumdar2,derrida1,derrida2,majumdar3,majumdar4}, diffusion models\cite{majumdar5,majumdar6},  reaction-diffusion systems\cite{cardy, fisher} to population dynamics\cite{frachebourg}. In most of these systems persistence probability decays algebraically with a non-trivial exponent $\lambda$ such as $P_{\zeta}(t) \sim t^{-\lambda}$. However, exponential and stretched exponential decays of  persistence probability has also been observed\cite{majumdar2}. Figure \ref{fig:persistence} 
shows the persistent probability of $\zeta$ at two different packing fractions and at various vision angles. We have found that in most cases the persistence probability does not decay algebraically or exponentially for the two packing fractions we studied. We have also observed that, in general, all the persistence probabilities can be fitted with a decaying stretched exponential function.  So the persistence probability in Figure \ref{fig:persistence} are fitted with a stretched exponential function $\sim e^{-(t/\tau)^\beta}$.  In Figure \ref{fig:tau_beta}, we have plotted $\beta$ and $\tau$ against the vision angles for all the two packing fractions.  It is clear from the values of $\beta$ that the persistence probability deviates from exponential decay. For the low packing fraction,  $\beta$ is very close to one for lower vision angles and then starts decreasing at $\theta$  larger than $30^\circ$  indicating the slowing down in randomising the motion.  However, at mid vision angles ($45^\circ$ to $60^\circ$), the value of $\beta$ increases above one. This indicate that the persistence probability decays faster than exponential, indicating that the direction of particle motion reverses much faster in this range of vision angles. For larger vision angles,  $\beta$ becomes smaller than unity, indicating that again  the direction of motion of the particles persists for a much larger time.  This is the range of vision angles where rotating clusters emerge.  So the persistent motion for the particles which are part of these rotating clusters  is short lived as they have nearly circular trajectories (Figure \ref{fig:trajectory}(c)). This is reflected in the persistence probability where the stretched exponential exponent $\beta$ takes a value larger than unity and goes through a maximum.  For even larger vision angles, $\beta$, again goes below unity indicating long persistent motion. In this range of vision angles, the system has larger aggregates, however they are not rotating.  So the particles in the aggregates move together and have longer persistent paths. For larger packing fractions,  the exponent decreases from  the values near unity at even lower vision angles. This can be understood from the fact that the in these packing fractions, the aggregate starts forming at  a lower range of vision angles and particles in these aggregates move together, resulting in a longer persistent motion. However, the value of $\beta$ increases again above unity for the mid range of vision angles as seen by the peak in $\beta$.  The time constant of this persistent probability $\tau$ systematically increases with  increase in vision angles, which is consistent with larger structures forming.

\section{Conclusions}

 We have carried out Langevin dynamics simulations of active Brownian particles having a visual perception of their immediate neighbours and adjust their activity accordingly.   Here each active particle is equipped with a vision cone and the presence of neighbours in this vision cone affects the rotational motion of the active particles.  We have observed that various structure emerge as we increase the vision angle of the active particles ranging from dilute fluids to large hexagonal clusters. It has been observed that  when the vision angle is in the intermediate range, these hexagonal clusters starts rotating spontaneously. We have characterised the rotational motion of these clusters using orientational autocorrelation function,  average angular momentum and average cluster size.  At low densities, largest cluster sizes are found at intermediate range of vision angles and also they posses the largest angular momentum. Thus the cluster size plays an important role in the rotation of aggregates. Earlier, rotating clusters are observed in anisotropic active particles. However in our model system, anisotropy is introduced by the vision cone in the self-propulsion direction of active particles.   It has also been observed that the emergent structures at different vision angles and their subsequent dynamics plays a bigger role in the persistence of dynamics of these iABPs.  We have found that the persistent motion of these particles is short lived in the mid range of vision angles and this could be correlated to the formation of larger aggregates  and their rotational motion as a whole.   Our results indicate that the intermediate range of vision angles is quite special in the structure formation and dynamics of the iABPs. This  is quite intriguing and further analysis is needed to unravel the physical reasons behind this spontaneous rotation of clusters in the absence of external torque.

\begin{acknowledgments}
	The authors acknowledge financial support from the Department of Atomic Energy, India 
	through the plan project (RIN4001-SPS).
\end{acknowledgments}

\section*{Supplementary Information}
The supplementary information~\cite{suppinfo}  includes GIFs for higher packing fraction demonstrating clusters that are rotating at an intermediate angle (M1.gif) and showing no significant rotation at a higher angle (M2.gif).

%\bibliography{qsps} 

\begin{thebibliography}{99}
\bibitem{ramaswamy1} S.~Ramaswamy, {\it Annu. Rev. Condens. Matter Phys.}, {\bf 1}, 323 (2010).
\bibitem{vicsek1} T.~Vicsek and A.~Zafeiris, {\it Phys. Rep.}, {\bf 517}, 71 (2012).
\bibitem{elgeti} J.~Elgeti, R.~G.~Winkler and G.~Gompper, {\it Rep. Prog. Phys.}, {\bf 78}, 056601 (2015).
\bibitem{reynolds} C.~W.~Reynolds, {\it Proc. 14th Annu. Conf. Comput. Graphics and Interactive Techniques}, 1987, pp.~25–34.
\bibitem{vicsek} T.~Vicsek, A.~Czirok, E.~Ben-Jacob, I.~Cohen, and O.~Shochet, {\it Phys. Rev. Lett.}, {\bf 75}, 1226 (1995).
\bibitem{couzin} I.~D.~Couzin, J.~Krause, R.~James, G.~D.~Ruxton and N.~R.~Franks, {\it J. Theor. Biol.}, {\bf 218}, 1 (2002).
\bibitem{couzin1} I.~D.~Couzin, J.~Krause, N.~R.~Franks and S.~A.~Levin, {\it Nature}, {\bf 433}, 513 (2005).
\bibitem{toner} J.~Toner and Y.~Tu, {\it Phys. Rev. E}, {\bf 58}, 4828 (1998).
\bibitem{mermin} N.~D.~Mermin and H.~Wagner, {\it Phys. Rev. Lett.}, {\bf 17}, 1133 (1966).
\bibitem{hohenberg} P.~Hohenberg, {\it Phys. Rev.}, {\bf 158}, 383 (1967).
\bibitem{ramaswamy} S.~Ramaswamy, R.~A.~Simha and J.~Toner, {\it Europhys. Lett.}, {\bf 62}, 196 (2003).
\bibitem{narayan} V.~Narayan, S.~Ramaswamy and N.~Menon, {\it Science}, {\bf 317}, 105 (2007).
\bibitem{tailleur} J.~Tailleur and M.~E.~Cates, {\it Phys. Rev. Lett.}, {\bf 100}, 218103 (2008).
\bibitem{redner} G.~S.~Redner, M.~F.~Hagan and A.~Bhaskaran, {\it Phys. Rev. Lett.}, {\bf 110}, 055701 (2013).
\bibitem{filly} Y.~Filli and M.~C.~Marchetti, {\it Phys. Rev. Lett.}, {\bf 100}, 218103 (2008).
\bibitem{cates} M.~E.~Cates and J.~Tailleur, {\it Annu. Rev. Condens. Matter Phys.}, {\bf 6}, 219 (2015).
\bibitem{grossman} D.~Grossman, I.~S.~Aranson and E.~B.~Jacob, {\it New J. Phys.}, {\bf 10}, 023036 (2008).
\bibitem{strombom} D.~Strombom, {\it J. Theor. Biol.}, {\bf 283}, 145 (2011).
\bibitem{szabo} B.~Szabo, G.~J.~Szollosi, B.~Gonci, Z.~Juranyi, D.~Selmeczi and T.~Vicsek, {\it Phys. Rev. E}, {\bf 74}, 061908 (2006).
\bibitem{peruani} F.~Peruani, A.~Deutsch and M.~Bar, {\it Phys. Rev. E}, {\bf 74}, 030904 (2006).
\bibitem{elgeti1} J.~Elgeti and G.~Gompper, {\it Europhys. Lett.}, {\bf 101}, 48003 (2013).
\bibitem{howse} J.~R.~Howse, R.~A.~Jones, A.~J.~Ryan, T.~Gough, R.~Vafabakhsh, and R.~Golestanian, {\it Phys. Rev. Lett.}, {\bf 99}, 048102 (2007).
\bibitem{erbe} A.~Erbe, M.~Zientara, L.~Baraban, C.~Kreidler, and P.~Leiderer, {\it J. Phys.: Condens. Matter}, {\bf 20}, 404215 (2008).
\bibitem{palacci} J.~Palacci, C.~Cottin-Bizonne, C.~Ybert, and L.~Bocquet, {\it Phys. Rev. Lett.}, {\bf 105}, 088304 (2010).
\bibitem{baraban} L.~Baraban, M.~Tasinkevych, M.~N.~Popescu, S.~Sanchez, S.~Dietrich, and O.~Schmidt, {\it Soft Matter}, {\bf 8}, 48–52 (2012).
\bibitem{dreyfus} R.~Dreyfus, J.~Baudry, M.~L.~Roper, M.~Fermigier, H.~A.~Stone and J.~Bibette, {\it Nature}, {\bf 437}, 862 (2005).
\bibitem{wilson} D.~A.~Wilson, R.~J.~Nolte and J.~C.~Van Hest, {\it Nature Chem.}, {\bf 4}, 268 (2012).
\bibitem{mukherjee} S.~Mukherjee and B.~L.~Bassler, {\it Nat. Rev. Microbiol.}, {\bf 17}, 371 (2019).
\bibitem{miller} M.~B.~Miller and B.~L.~Bassler, {\it Annu. Rev. Microbiol.}, {\bf 55}, 165 (2001).
\bibitem{jalim} J.~Singh and A.~V.~A.~Kumar, {\it Phys. Rev. E}, {\bf 101}, 022606 (2020).
\bibitem{bauerle} T.~Bauerle, A.~Fischer, T.~Speck and C.~Bechinger, {\it Nat. Commun.}, {\bf 9}, 3232 (2018).
\bibitem{cao} J.~Cao, J.~Wu and Z.~Hou, {\it Phys. Chem. Chem. Phys.}, {\bf 26}, 7783 (2024).
\bibitem{lavergne} F.~A.~Lavergne, H.~Wendehenne, T.~Bauerle, and C.~Bechinger, {\it Science}, {\bf 364}, 70 (2019).
\bibitem{barberis} L.~Barberis and F.~Peruani, {\it Phys. Rev. Lett.}, {\bf 117}, 248001 (2016).
\bibitem{negi} R.~S.~Negi, R.~G.~Winkler and G.~Gompper, {\it Soft Matter}, {\bf 18}, 6167 (2022).
\bibitem{yang} Y.~Yang, F.~Qiu and G.~Gompper, {\it Phys. Rev. E}, {\bf 89}, 012720 (2014).
\bibitem{huang} D.~Huang, Y.~Du, H.~Jiang and Z.~Hou, {\it Phys. Rev. E}, {\bf 104}, 034606 (2021).
\bibitem{cantisan} J.~Cantisan, J.~M.~Seoane and M.~A.~F.~Sanjuan, {\it Chaos, Solitons and Fractals}, {\bf 172}, 113531 (2023).
\bibitem{kokot} G.~Kokot and A.~Snezhko, {\it Nature Commun.}, {\bf 9}, 1 (2018).
\bibitem{ciccotti} E.~Vanden-Eijnden and G.~Ciccotti, {\it Chem. Phys. Lett.} {\bf 429}, 310 (2006).
\bibitem{EM} P.~E.~Kloeden and E.~Platen, {\it Numerical Solution of Stochastic Differential Equations}, Springer, Berlin (1992). ISBN: 3-540-54062-8.
\bibitem{mandal} S.~Mandal, B.~Liebchen and H.~Lowen, {\it Phys. Rev. Lett.}, {\bf 123}, 228001 (2019).
\bibitem{grzybowski} B.~A.~Grzybowski, H.~A.~Stone and G.~M.~Whitesides, {\it Nature}, {\bf 405}, 1033 (2000).
\bibitem{yan} J.~Yan, M.~Bloom, S.~C.~Bae, E.~Luijten and S.~Granick, {\it Nature}, {\bf 491}, 578 (2012).
\bibitem{grier} D.~G.~Grier, {\it Curr. Opin. Colloid Interface Sci.}, {\bf 2}, 264 (1997).
\bibitem{moffitt} J.~R.~Moffitt, Y.~R.~Chemla, S.~B.~Smith and C.~Bustamante, {\it Annu. Rev. Biochem.}, {\bf 77}, 205 (2008).
\bibitem{drescher} K.~Drescher, K.~C.~Leptos, I.~Tuval, T.~Ishikawa, T.~J.~Pedley and R.~E.~Goldstein, {\it Phys. Rev. Lett.}, {\bf 102}, 168101 (2009).
\bibitem{riedel} I.~H.~Riedel, K.~Kruse and J.~Howard, {\it Science}, {\bf 309}, 300 (2005).
\bibitem{kummel} F.~Kummel, B.~ten~Hagen, R.~Wittkowski, I.~Buttinoni, R.~Eichhorn, G.~Volpe, H.~Lowen and C.~Bechinger, {\it Phys. Rev. Lett.}, {\bf 110}, 198302 (2013).
\bibitem{vaccari} L.~Vaccari, M.~Molaei, R.~L.~Leheny and K.~J.~Stebe, {\it Soft Matter}, {\bf 14}, 5643 (2018).
\bibitem{majumdar1} S.~N.~Majumdar, {\it Curr. Sci.}, {\bf 77}, 370 (1999).
\bibitem{majumdar2} S.~N.~Majumdar and A.~J.~Bray, {\it Phys. Rev. Lett.}, {\bf 91}, 030602 (2003).
\bibitem{derrida1} B.~Derrida, A.~J.~Bray and C.~Godreche, {\it J. Phys. A}, {\bf 27}, L357 (1994).
\bibitem{derrida2} B.~Derrida, B.~Hakkim and V.~Paqsuier, {\it Phys. Rev. Lett.}, {\bf 75}, 751 (1995).
\bibitem{majumdar3} S.~N.~Majumdar and C.~Sire, {\it Phys. Rev. Lett.}, {\bf 77}, 1420 (1996).
\bibitem{majumdar4} S.~N.~Majumdar, A.~J.~Bray, S.~J.~Cornell and C.~Sire, {\it Phys. Rev. Lett.}, {\bf 77}, 3704 (1996).
\bibitem{majumdar5} S.~N.~Majumdar, C.~Sire, A.~J.~Bray and S.~J.~Cornell, {\it Phys. Rev. Lett.}, {\bf 77}, 2867 (1996).
\bibitem{majumdar6} S.~N.~Majumdar and A.~J.~Bray, {\it Phys. Rev. Lett.}, {\bf 81}, 2626 (1998).
\bibitem{cardy} J.~Cardy, {\it J. Phys. A}, {\bf 28}, L19 (1995).
\bibitem{fisher} D.~S.~Fisher, P.~Le~Doussal and C.~Monthus, {\it Phys. Rev. Lett.}, {\bf 80}, 3539 (1998).
\bibitem{frachebourg} L.~Frachebourg, P.~L.~Krapivsky and E.~Ben-Naim, {\it Phys. Rev. Lett.}, {\bf 77}, 2125 (1996).
\bibitem{suppinfo} Supplementary Information for this work can be accessed at [link to supplementary material].

\end{thebibliography}
\end{document}